\documentclass[twocolumn,twocolappendix]{aastex631}
\shortauthors{Chen et al.}
\shorttitle{}

\begin{document}


\title{The circumgalactic medium traced by Mg\,$\textsc{II}$ absorption with DESI: dependence on galaxy stellar mass, star formation rate and azimuthal angle}

\author[0009-0004-5989-6005]{Zeyu Chen}\thanks{Email:  \href{mailto:czy664@mail.ustc.edu.cn}{czy664@mail.ustc.edu.cn}}
\affiliation{Department of Astronomy, University of Science and Technology of China, Hefei 230026, China}
\affiliation{School of Astronomy and Space Science, University of Science and Technology of China, Hefei 230026, China}

\author[0000-0003-1588-9394]{Enci Wang}\thanks{Email:  \href{mailto:ecwang16@ustc.edu.cn}{ecwang16@ustc.edu.cn}}
\affiliation{Department of Astronomy, University of Science and Technology of China, Hefei 230026, China}
\affiliation{School of Astronomy and Space Science, University of Science and Technology of China, Hefei 230026, China}

\author[0000-0002-6684-3997]{Hu Zou}
\affiliation{National Astronomical Observatories, Chinese Academy of Sciences, Beijing 100012, China}

\author[0000-0002-3983-6484]{Siwei Zou}
\affiliation{Chinese Academy of Sciences South America Center for Astronomy, National Astronomical Observatories, CAS, Beijing 100101, China}
\affiliation{Department of Astronomy, Tsinghua University, Beijing 100084, China}

\author{Yang Gao}
\affiliation{College of Physics and Electronic Information, Dezhou University, Dezhou 253023, China}

\author[0000-0002-4911-6990]{Huiyuan Wang}
\affiliation{Department of Astronomy, University of Science and Technology of China, Hefei 230026, China}
\affiliation{School of Astronomy and Space Science, University of Science and Technology of China, Hefei 230026, China}

\author[0009-0008-1319-498X]{Haoran Yu}
\affiliation{Department of Astronomy, University of Science and Technology of China, Hefei 230026, China}
\affiliation{School of Astronomy and Space Science, University of Science and Technology of China, Hefei 230026, China}

\author[0009-0004-7042-4172]{Cheng Jia}
\affiliation{Department of Astronomy, University of Science and Technology of China, Hefei 230026, China}
\affiliation{School of Astronomy and Space Science, University of Science and Technology of China, Hefei 230026, China}

\author[0009-0009-2660-1764]{Haixin Li}
\affiliation{Department of Astronomy, University of Science and Technology of China, Hefei 230026, China}
\affiliation{School of Astronomy and Space Science, University of Science and Technology of China, Hefei 230026, China}

\author[0009-0006-7343-8013]{Chengyu Ma}
\affiliation{Department of Astronomy, University of Science and Technology of China, Hefei 230026, China}
\affiliation{School of Astronomy and Space Science, University of Science and Technology of China, Hefei 230026, China}

\author[0000-0002-6873-8779]{Yao Yao}
\affiliation{Department of Astronomy, University of Science and Technology of China, Hefei 230026, China}
\affiliation{School of Astronomy and Space Science, University of Science and Technology of China, Hefei 230026, China}

\author[0000-0003-0230-4596]{Weiyu Ding}
\affiliation{Department of Astronomy, University of Science and Technology of China, Hefei 230026, China}
\affiliation{School of Astronomy and Space Science, University of Science and Technology of China, Hefei 230026, China}
\affiliation{National Astronomical Observatories, Chinese Academy of Sciences, Beijing 100012, China}

\author[0000-0002-8037-3573]{Runyu Zhu}
\affiliation{National Astronomical Observatories, Chinese Academy of Sciences, Beijing 100012, China}
\affiliation{Department of Astronomy, Xiamen University, Xiamen, Fujian 361005, China}




\begin{abstract}

Understanding the circumgalactic medium (CGM) distribution of galaxies is the key to revealing the dynamical exchange of materials between galaxies and their surroundings. In this work, we use DESI EDR dataset to investigate the cool CGM of galaxies ($0.3<z<1.7$) with stacking the spectra of background QSOs to obtain Mg\,\textsc{ii} absorption of foreground galaxies. The equivalent width of Mg\,\textsc{ii} absorption strongly correlates to stellar mass with \rm{EW(Mg\,\textsc{ii})} $\propto M_{*}^{0.5}$ for star-forming galaxies with $\log M_{*}/M_{\odot} < 10$, but is independent with mass for galaxies above this mass. At given stellar mass, \rm{EW(Mg\,\textsc{ii})} is larger for galaxies of higher star formation rate with impact parameter less than $50$ kpc, while showing little dependence on galaxy size. By studying the dependence on azimuthal angle, we find \rm{EW(Mg\,\textsc{ii})} is strongest at the direction near the minor axis for star-forming galaxies with $\log M_{*}/M_{\odot} < 10.0$, while no dependence on azimuthal angle is seen for luminous red galaxies. This indicates that the outflow associated with star formation enhances the Mg\,\textsc{ii} absorption. However, for galaxies with $\log M_{*}/M_{\odot} > 10.0$,  the \rm{EW(Mg\,\textsc{ii})} at the minor axis is largely suppressed with respect to low mass galaxies. This suggests that the competing processes, such as stellar feedback and gravity, play a key role in shaping the distribution of outflowing gas.

\end{abstract}



\section{Introduction} \label{sec:intro}

The circumgalactic medium (CGM) surrounding galaxies plays a crucial role in driving the formation and evolution of galaxies \citep[e.g.][]{tumlinson17,Peroux-20}. In recent years, simulations and observations of this component of galaxies are progressing rapidly \citep[e.g.][]{nelson20,nelson23,lim21,defelippis24}. High-resolution ``zoom-in" simulations of {\tt Eris2} have demonstrated that galaxies can produce large-scale galactic outflows, which is crucial for generating realistic dwarf galaxies and large late-type spiral galaxies \citep{governato10,guedes11}. \cite{shen13} found that about one-third of all the gas within $R_{\rm vir}$ is outflowing and inflows bring in material along optically thick ``cold'' streams in their simulation  of $z \sim 3$ massive galaxies.  
More recently, there are increasing evidences in simulations that the inflowing gas is almost co-planar and more or less co-rotating with the gas disk \citep[e.g.][]{ Danovich-15, Stewart-17, Peroux-20, Trapp-21, Hafen-22, Gurvich-22,faucher2023}, while the outflowing gas driven by stellar feedback is preferentially along the direction that is perpendicular to the disk \citep{Nelson-19, Peroux-20, SenWang22, weng24}.

Observationally, the technique of absorption line spectroscopy is a powerful tool to extract the properties of CGM. By observing the spectra of distant quasi-stellar objects (QSOs), we can detect absorption features within the CGM of these foreground galaxies, thereby obtaining valuable information about the surrounding galactic environment. By stacking the spectra of background QSO at the redshift of foreground galaxies, \cite{zhu13ca} studied the Ca\,\textsc{ii} H and K line absorption of galaxies at $z \sim 0.1$, and discovered that bipolar outflows induced by star formation must have played a significant role in producing Ca\,\textsc{ii} in galaxy halos. \cite{liang14} found differential covering fraction between low- and high-ionization gas based on observations of \text{Ly}\ensuremath{\alpha}, C\,\textsc{ii}, C IV, Si\,\textsc{ii}, Si\,\textsc{iii}, and Si\,\textsc{iv} absorption transitions. 
More recently, \cite{Zou-24b} detected nine high-redshift ($z > 6$) absorbers in quasar spectra, including O\,\textsc{i}, Mg\,\textsc{ii}, C\,\textsc{ii}, C\,\textsc{iv}, Si\,\textsc{ii} and Fe\,\textsc{ii} , and found that the absorbing gas with a higher [$\alpha$/Fe] is possibly associated with a less overdense region.

Among all the detected metal lines, the most extensively examined is the $\rm{Mg\,\textsc{ii}} \thinspace \lambda\lambda 2796, 2803$ doublet.  \cite{bergeron86} first discovered the Mg\,\textsc{ii} doublet absorption lines in a pair of galaxies, and claimed that this metal absorber is a crucial indicator of enriched cold gas. The Mg\,\textsc{ii} absorption originally arises from photo-ionized gas with temperatures of approximately $T\sim$ 10$^{4} K$  \citep{bergeron86,charlton03} at the neutral Hydrogen column densities ranging from ${\rm N(H\,\textsc{i})} \approx$ 10$^{16}$ -- 10$^{22}$ cm$^{-2}$  \citep{churchill00,rao06}. Mg\,\textsc{ii} absorption is ubiquitously detected in the gaseous halos in and around galaxies and has become an effective tool for studying the distribution of cool gas in the CGM. The Voigt profile fitting of the Mg II doublet also provides information about the absorber’s column density and Doppler broadening \citep[e.g.][]{wu2023}.

By associating the Mg\,\textsc{ii} absorption systems with the locations of galaxies, \cite{dutta20} obtained the absorption strength of the CGM of galaxies at different impact parameters\footnote{The impact parameter is defined as the projected distance between the absorber to the galaxy at the galaxy redshift.} ($R$), and found a trend of stronger Mg\,\textsc{ii} absorption being associated with more massive galaxies, and to a lesser extent, with higher star forming galaxies.  
\cite{anand21} calculated the excess surface density of Mg\,\textsc{ii} absorption systems around galaxies of different types and found there is a weak evolution of \rm{EW(Mg\,\textsc{ii})} with galaxy redshift and emission line galaxies (ELGs) have 2–4 times higher covering fractions than luminous red
galaxies (LRGs) within $R \thinspace [\rm{kpc}]<50$.  
%
Using strong intervening absorption systems in the near-IR spectra of 31 luminous quasars, \cite{Zou-21} found the line density 
of strong Mg\,\textsc{ii} absorbers decrease toward higher redshifts at $z>3$. Using deep JWST/NIRCam slitless grism spectroscopy of quasar J0100+2802, \cite{bordoloi24} discovered the Mg\,\textsc{ii} radial absorption profile falls off sharply with radius,  extending out to 2–3 $R_{200}$ of the host galaxies.   

However, the study of individual absorbers usually requires high signal-to-noise spectra and strong absorption features \citep[e.g.][]{Chen-17, Zou-21, Davies-23, napolitano23, bordoloi24, Sebastian-24, Zou-24a}, which may be biased to trace dense and metal-rich gas in CGM. 
This therefore calls for the spectra stacking technique to obtain the mean or median absorption features of CGM gas at given impact parameters.  
By stacking the SDSS spectra, \cite{menard11} showed that Mg\,\textsc{ii} absorbers trace a substantial fraction of the [O\,\textsc{ii}] luminosity density in the host galaxy \citep[also see][]{Joshi-17}.  And by using co-added spectra of more than 5000 background galaxies from zCOSMOS Redshift Survey \citep{lilly07}, \cite{bordoloi11} found a clear correlation between Mg\,\textsc{ii} absorption line strength and the host galaxy stellar mass for the blue galaxies.  Later \cite{bordoloi14} found that the blue galaxies are associated with a much stronger outflowing component as compared to the red galaxies in terms of their rest frame equivalent widths \citep[also see][]{lan18}. 



In this paper, we aim at revisiting the Mg\,\textsc{ii} absorption features in the CGM of foreground galaxies from the background QSOs, using the largest on-going spectroscopy survey, Dark Energy Spectroscopic Instrument \citep[DESI;][]{aghamousa16a,aghamousa16b}.  
DESI aims at mapping the nature of dark energy with spectroscopic measurements of 40 million galaxies and quasars. 
Compared to other notable large programs such as MUSE Analysis of Gas around Galaxies (MAGG; \cite{lofthouse2020,dutta20,fossati2021}), MusE GAs FLOw and Wind (MAGAFLOW; \cite{schroetter2016,zabl2019,schroetter19,zabl2020}), The Mg\,\textsc{ii} Absorber–Galaxy Catalog (MAGIICAT; \cite{nielsen13,nielsen13b,churchill2013,nielsen2016}), and MUSE-ALMA Haloes Survey \citep{hamanowicz2020,szakacs2021,peroux2022}, DESI has unique advantages in terms of the volume of data that will be released in the future. The DESI Early Data Release has released 437,664 ELGs,  227,318 LRGs, and 76,079 QSOs. 
These large mount spectra enable us to explore the Mg\,\textsc{ii} absorption  with stacking technique to unprecedented details.   As the first paper of this series,  we mainly investigate the dependence of the \rm{EW(Mg\,\textsc{ii})} on a set of parameters, such as stellar mass ($M_*$), star formation rate (SFR),  half-light radius ($R_{\rm e}$) and inclination of host galaxies.   We also investigate \rm{EW(Mg\,\textsc{ii})} at different impact parameters ($R$) as well as different azimuthal angles from the host galaxies.   

This paper is structured as follows.  We  introduce the DESI survey,  the galaxy sample of DESI EDR,  and the spectral stacking technique in Section \ref{sec:analysis}. In Section \ref{subsec:RMdependence}, we focus on discussing the impact of stellar mass on the \rm{EW(Mg\,\textsc{ii})} for ELGs and the variation of \rm{EW(Mg\,\textsc{ii})} with impact parameter. In Section \ref{subsec:redshift}, we examine the  evolution of Mg\,\textsc{ii} absorption strength, with controlling stellar mass of galaxies.  We also investigate the influence of SFR and effective radius on the \rm{EW(Mg\,\textsc{ii})}, as well as the angular distribution in the CGM, in Section \ref{subsec:SFRandRe} and \ref{subsec:angledependence}.  In Section \ref{sec:discussion}, we discuss the covering fraction of Mg\,\textsc{ii} in galaxies and the mass integral of H\,\textsc{i} in CGM.  Finally, we summarize our results  in Section \ref{sec:summary}. We adopt a flat $\Lambda$CDM cosmology with $h=0.677$,  $\Omega_{\rm M}=0.309$ and $\Omega_{\rm Lambda}=0.691$ using the \href{https://docs.astropy.org/en/latest/api/astropy.cosmology.realizations.Planck15.html}{\tt Planck15} package \citep{2016A&A...594A..13P}.

\section{DATA analysis} \label{sec:analysis}
In this work, we perform a positional cross-matching between QSOs and galaxies in DESI EDR. The QSO-galaxy pairs we investigated are mostly within 200 kpc, which results in roughly 10,000 QSO-ELG pairs and 6,000 QSO-LRG pairs. In the work, we do not exclude the duplicated observations of QSOs on different dates, but treat them as different observational samplings in the stacking technique.  We then stack the spectra of the background QSOs at redshift of foreground galaxies to obtain a high signal-to-noise ratio stacked spectrum, from which we extracted statistical information about the CGM of the foreground galaxies. The detailed steps will be introduced in Section \ref{subsec:method}. 

\subsection{DESI EDR} \label{subsec:DESI}
DESI Survey \citep{aghamousa16a,aghamousa16b} is the largest multiobject spectrograph survey constructed to date and was designed to efficiently conduct a comprehensive spectroscopic exploration into the nature of dark energy \citep{abareshi22}.   As one of the latest and most important dark energy survey projects, DESI has unique advantages in terms of observational scale, depth, and efficiency. 
DESI's large survey in a short time is enabled by substantial instrument development, allowing it to record up to 5,000 spectra in a single exposure, with high throughout and operational efficiency \citep{levi19}. A comprehensive description of the completed instrument can be found in \cite{abareshi22}.

Being the first public release of DESI survey, the DESI \href{https://data.desi.lbl.gov/doc/releases/edr/}{Early Data Release (EDR)} includes the spectra for 1.8 million unique targets from Survey Validation observations taken from December 2020 through June 2021, as well as a few commissioning and special tiles observed during the same time period \citep{adame23}. 
The EDR comes from the ``Survey Validation" (SV) phase prior to the main survey, aiming to refine target selection and observing procedures \citep{adame23,myers23}. The initial SV1 used relative looser cuts (see \cite{guy2023}) to build truth samples and optimize signal-to-noise requirements. This was followed by operation phase (SV2) and a ``One-Percent Survey" (SV3) that further improved observing efficiency and fiber assignment completeness over a small area.


\subsection{Datasets} \label{subsec:dataset}
Galaxies in DESI EDR can be categorized into three classes \citep{myers23}: Emission Line Galaxies \citep{raichoor20,raichoor23}, Luminous Red Galaxies  \citep{zhou20,zhou23}, and Bright Galaxy Survey (BGS) \citep{ruiz20,hahn23}. We utilize the {\tt{SPECTYPE}} from the {\tt  \href{https://data.desi.lbl.gov/public/edr/spectro/redux/fuji/zcatalog/zall-tilecumulative-fuji.fits}{zcatalog} } of  DESI project to distinguish between QSOs \citep{yeche20,chaussidon23} and galaxies.  

The redshift estimation algorithm for galaxies in this catalog is known as \href{https://github.com/desihub/redrock}{\tt redrock}.  The stellar masses of galaxies are also taken from \href{https://data.desi.lbl.gov/public/edr/vac/edr/stellar-mass-emline/v1.0/edr_galaxy_stellarmass_lineinfo_v1.0.fits}{the EDR value-added catalog}, which are estimated using the CIGALE SED fitting code \citep{boquien19}. The continuum SNR especially for ELGs is low, which maks it hard to constrain the stellar mass by only using full spectral fitting techniques. So the SED  fitting procedure combines broad-band photometry (g, r, z, W1, W2) from the DESI Legacy Imaging Surveys and spectrophotometry of 10 synthetic bands derived from DESI spectra. These 10 synthetic bands are obtained by convolving the optical spectra with self-defined 10 contiguous broad bands, which start from 3,650 Å and have bandwidths of about 615 Å (see \cite{zou24}).
The main optical emission lines are measured by single Gaussian fitting without continuum modeling \citep{zou24}. The ELGs/LRGs in DESI EDR have a typical stellar mass of about $10^{9.81}M_{\odot}$/$10^{11.31}M_{\odot}$. 
We also adopt the half-light radius, ellipticity component 1 ($\epsilon_1$), and ellipticity component 2 ($\epsilon_2$) from this catalog. We convert them into the minor-to-major axis ratio (b/a) and position angle ($\rm{\theta}$) of galaxies using the methods provided on the \href{https://www.legacysurvey.org/dr10/catalogs/}{DESI Legacy Survey website}, as shown in the following Equation \ref{eq:epsilon}, \ref{eq:b/a} and \ref{eq:theta}.  

\begin{equation}
\epsilon=\frac{a-b}{a+b}\exp(2i\theta)=\epsilon_1+i\epsilon_2;
\label{eq:epsilon}
\end{equation}

\begin{equation}
\frac{b}{a}=\frac{1-|\epsilon|}{1+|\epsilon|};
\label{eq:b/a}
\end{equation}

\begin{equation}
\theta = \frac{1}{2}\arctan\frac{\epsilon_2}{\epsilon_1}.
\label{eq:theta}
\end{equation}

We use the luminosity of [O\,\textsc{ii}] $\lambda\lambda3726,3729$ doublet  to estimate the SFR of galaxies, following the formula \citep{schroetter19}: 
\begin{equation}
\text{SFR}([\text{O\,\textsc{ii}}]) = 4.1 \times 10^{-42} (\text{L}\text{[O\,\textsc{ii}]} \thinspace\text{erg} \thinspace\text{s}^{-1} ) \thinspace\text{M}_{\odot} \thinspace\text{yr}^{-1}.
\label{eq:SFROII}
\end{equation}
To correct both the Milky Way \citep{schlegel98} and intrinsic galaxy extinction, we adopt the dust extinction law of \cite{calzetti00}.   
Following the approach of \cite{schroetter19}, we calculate the ${\rm E(B - V)}$ of DESI galaxies employing the $M_{*} - {\rm E(B - V)}$ relation for \cite{chabrier03} initial mass function established by \cite{garn10},  which can be written as:  
\begin{equation}
{\rm E(B - V)} = (0.93 + 0.77X + 0.11X^2 - 0.09X^3)/k_{\rm H\alpha}, 
\label{eq:ebv}
\end{equation}
where $X = \log(M_{*}/{\rm M_\sun}) - 10$ and $k_{\rm H\alpha} = 3.326$ for the \cite{calzetti00} extinction law. 

\begin{figure}[htbp]
    \quad
    \includegraphics[width=0.91\columnwidth]{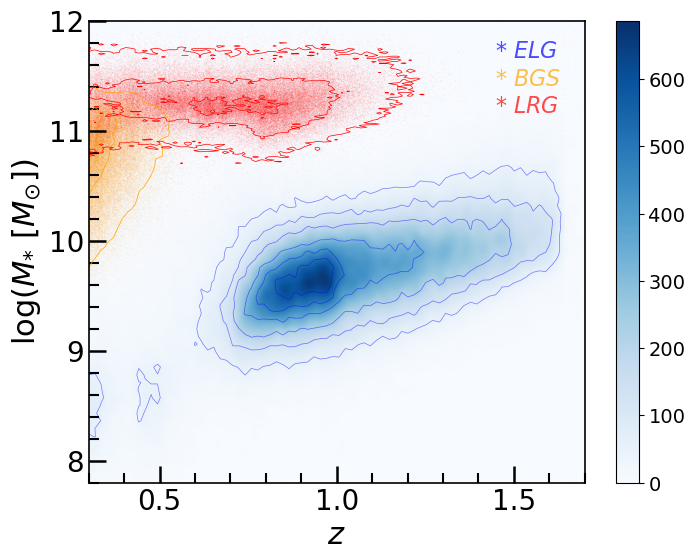}
    \par
    \quad\thinspace
    \includegraphics[width=0.927\columnwidth]{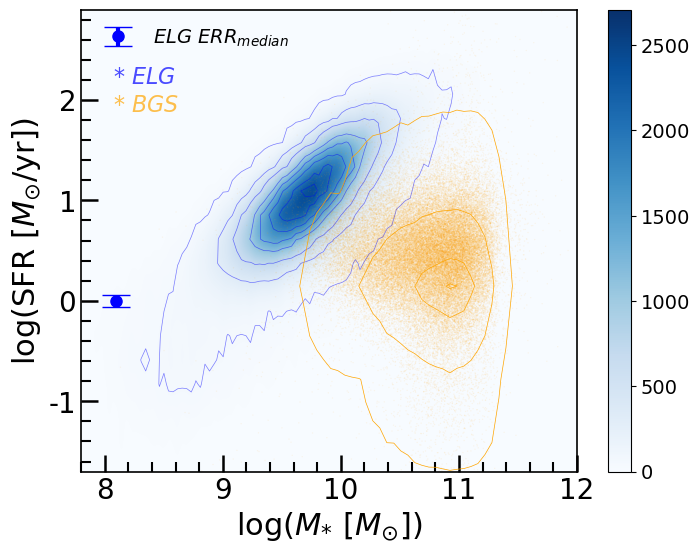}
    \caption{Top panel: redshift vs. stellar mass for galaxies in QSO-galaxy pairs with impact parameter less than 1200 kpc.  Bottom panel: SFR vs stellar mass for the same galaxies as above.  The blue erorrbar represents the median errors of SFR for ELGs. The solid lines represent their respective iso-density contours. The proportion of scattered points enclosed by each line forms an arithmetic sequence from 5\% to 95\%. The SFR for LRGs are not shown here due to the significantly larger uncertainties of [O\,\textsc{ii}] fluxes for LRGs in DESI compared to ELGs.}
    \label{fig:SFR}
\end{figure}

In Figure \ref{fig:SFR}, we show the stellar mass-SFR diagram and the mass-redshift diagram of galaxies that are in QSO-galaxy pairs with impact parameter less than 1200 kpc.  The BGS galaxies are located at low redshift ($z<0.5$), and includes both star-forming and quenched galaxies. The ELGs are located on a narrow range on stellar mass-SFR diagram, which is known as star formation main sequence \citep[SFMS;][]{brinchmann04,noeske07,daddi07,speagle14}. And LRGs are more massive than ELGs, but most are believed to be passive galaxies with low SFR \citep{wu2024}. The median uncertainty of SFR measurements, based on the uncertainties of [O\,\textsc{ii}] fluxes, is indicated with blue error bar in the bottom panel of Figure \ref{fig:SFR}. 
As shown in Figure \ref{fig:D1} of Appendix \ref{sec:appenD}, we compare the SFMS of DESI ELGs with those from previous works in the literature and find good consistency, validating the method of measuring SFRs.

The redshift range of ELGs is broader than that of LRGs, extending up to redshifts as high as 1.7, while the highest redshift of LRGs is approximately 1.2. We selected galaxies with $z \ge 0.3$ to ensure that the Mg\,\textsc{ii} absorption line wavelength is greater than the blue end of the DESI spectrum (3600\thinspace\AA).   

\subsection{Method} \label{subsec:method}

By matching the positions of QSOs and galaxies, we obtain all the QSO-galaxy pairs within an impact parameter of 1200 kpc.  In the matching, we require that the redshift of QSOs is higher than galaxies by at least 0.02, to avoid the potential contamination of absorption from QSOs themselves. According to \cite{napolitano23}, the velocity offset between the absorber and the QSO is defined as $v_{\text{off}} = c \frac{z_{\text{Mg II}} - z_{\text{QSO}}}{1 + z_{\text{QSO}}}$, and considering the redshift range of our foreground galaxies and the typical outflow velocity of QSOs, \(|z_{\text{Mg II}} - z_{\text{QSO}}| \geq 0.02\) is a relatively safe criterion for exclusion. The number density of QSO-galaxy pairs is proportional to the impact parameter,  since there is no expectation that the positions of background QSO correlate with the foreground galaxies.  In Figure \ref{fig:A1} of Appendix \ref{sec:appenA}, we have examined the impact parameter distribution of QSO-galaxy pairs, confirming the validity of our matching method.

Before the spectral stacking, the QSO spectra need to be well-processed to exclude the intrinsic features.  To do this, we first fit the QSO spectral with the \href{https://github.com/legolason/PyQSOFit}{fitting program} provided by \cite{guo18}.  
Then, we normalize the observed QSO spectra with the fitted ones, to remove global features. 
To further remove the small scale features (in wavelength), we applied a median filter with a window size of several pixels. We choose to adopt the same window size of 71 pixels as used by \cite{zhu13} for SDSS spectra, because the resolution of DESI spectra at 2800 Å is comparable to that of SDSS after redshifting ($z\sim0.8$). While doing so, we have masked the wavelength range containing Mg\,\textsc{ii} absorption lines to prevent its influence on the median filtering. In Appendix \ref{sec:appenE}, we show normalized spectra as well as our median filter and the spectra after applying the median filter.  After removing the small scale and large scale features of QSO spectra, the normalized spectra (shown in blue in Figure \ref{fig:spec} of Appendix \ref{sec:appenE}) are the ones that are used in spectral stacking. 

Finally,  we stack the normalized QSO spectra at the rest-frame wavelength of the foreground galaxies and take the arithmetic median of the flux at each wavelength point to obtain the co-added spectrum. We have also tested the results obtained using the $\sigma$-clipped mean and RMS noise-weighted mean stacking methods and found that the results from median stacking are comparable to those from the $\sigma$-clipped mean but exhibit lower noise. In fact, median stacking is widely favored \citep[e.g.][]{zhu13,bordoloi11,zhu13ca,bordoloi14,lan18} for its stability, resistance to outliers, suitability for low signal-to-noise ratio data, as well as its simplicity and reliability. Then we employ the bootstrap method to estimate the error at each wavelength with 100 times of sampling. 
For the stacked spectra, the wavelength pixel interval is set as 0.8 Å, which is the same as the DESI released spectra.  In appendix \ref{sec:appenC}, we have examined that the single-to-noise ratio of our stacked spectrum is proportional to the N$^{0.5}_{\rm{spec}}$ as expected, where N$_{\rm{spec}}$ is the number of spectra used in stacking.  

We perform a double-Gaussian fitting on the Mg\,\textsc{ii} absorption line component in the stacked  spectrum using {\tt \href{https://emcee.readthedocs.io/en/stable/index.html}{emcee}} \citep{foreman13}.  
For each individual stacked spectrum, we measure the \rm{EW(Mg\,\textsc{ii})} (the equivalent width of the $\rm{Mg\,\textsc{ii}} \thinspace \lambda\lambda 2796, 2803$ doublet) and its error with the following two approaches:  
\begin{itemize}
\item
One approach is to directly obtain the measurement of the equivalent width from the fitted Gaussian profile, and calculate its error using the models of MCMC at the final 5000 steps;

\item 
The other approach is to directly integrate the Mg\,\textsc{ii} absorption across the wavelengths that are located within  3-$\sigma$ of the fitted Gaussian model, and the root-mean-square (RMS) of the area errors for these wavelength points as the error of the equivalent width.

\end{itemize}

When the absorption signal is relatively strong, the results from the two approaches are almost identical.  In this work, we present the results with the second method,  because when the signal is weaker, the profile of Mg\,\textsc{ii} doublet may not  be well characterized as Gaussian functions. 
We have examined that using the measurements from Gaussian fittings does not change the main results of this work. 
Additionally, we also perform a visual inspection for each co-added spectrum. In case that the absorption signal is overwhelmed by the uncertainties in the co-added spectrum,  reliable measurements of  \rm{EW(Mg\,\textsc{ii})} are not possible.  We therefore take the RMS of the error values (i.e., 1-$\sigma$) over a fixed wavelength range that covers the Mg\,\textsc{ii} absorption line as an upper limit.
In Appendix \ref{sec:appenF}, we show some examples of co-added spectra, and the above two approaches to obtain \rm{EW(Mg\,\textsc{ii})}, as well as the case of only presenting the upper limit of \rm{EW(Mg\,\textsc{ii})} measurements. All the co-added spectra images used in this work are available on \href{https://docs.google.com/presentation/d/1jF6wtoihM-P7Dg6m74fkX0zbfMV2tt88/edit?usp=share_link&ouid=101552872223175249144&rtpof=true&sd=true}{our Google Drive}.



\section{RESULTS} \label{sec:result}

In this work, we mainly focus on the Mg\,\textsc{ii} absorption of ELGs, since ELGs are expected to be continuously accreting cool gas, forming stars and stripping gas away by stellar feedback all along their life time.  This active exchange of gas with surrounding environments may imprint on the distribution of CGM \cite[e.g.][]{tumlinson17}. 

\subsection{Mg\,\textsc{ii} absorption for galaxies of different stellar mass} \label{subsec:RMdependence}


\begin{figure*}
    \centering
    \includegraphics[width=1.9\columnwidth]{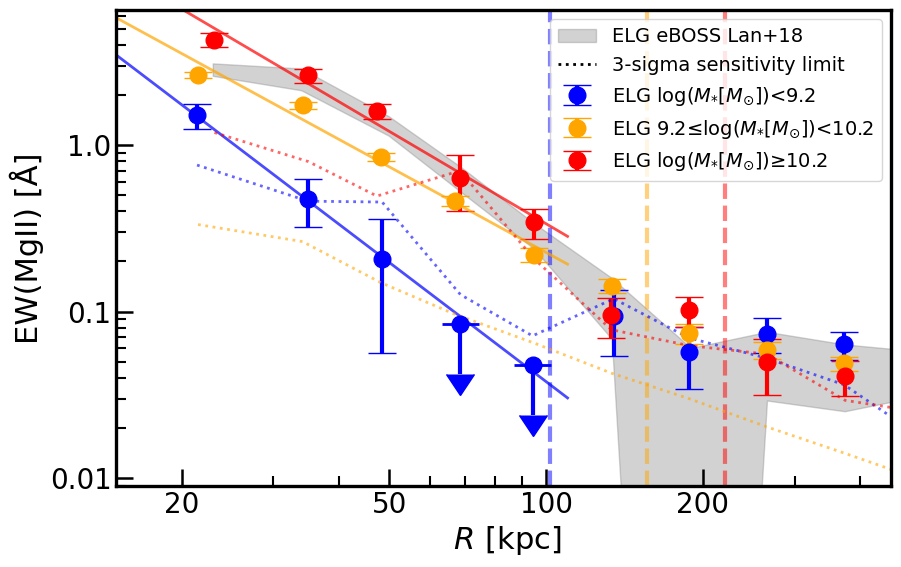}
    \caption{The \rm{EW(Mg\,\textsc{ii})} as a function of impact parameter for ELGs of different stellar masses.  The average stellar masses for the three mass bins are $10^{8.83}M_{\odot}$ (blue), $10^{9.76}M_{\odot}$ (orange) and $10^{10.51}M_{\odot}$ (red) in ascending order. The straight lines represent the corresponding linear fits at $R<100$ kpc, arranged in ascending order of stellar masses, which are $\log (\rm{EW [\text{\AA}]}) = -2.38\log(R [\rm{kpc}]) + 3.34$ (blue), $\log (\rm{EW [\text{\AA}]}) = -1.71\log(R [\rm{kpc}]) + 2.77$ (orange) and  $\log (\rm{EW [\text{\AA}]}) = -1.84\log(R [\rm{kpc}]) + 3.22$ (red), respectively. The dotted line represents the 3-sigma sensitivity limit of the spectra, and the dashed lines represent the average virial radius ($R_{\rm{vir}}$) of galaxies in three different stellar mass bins. The shaded gray area represents the results obtained by \cite{lan18} using eBOSS DR14.}
    \label{fig:elgew}
\end{figure*}

\begin{figure*}
    \centering
    \includegraphics[width=2\columnwidth]{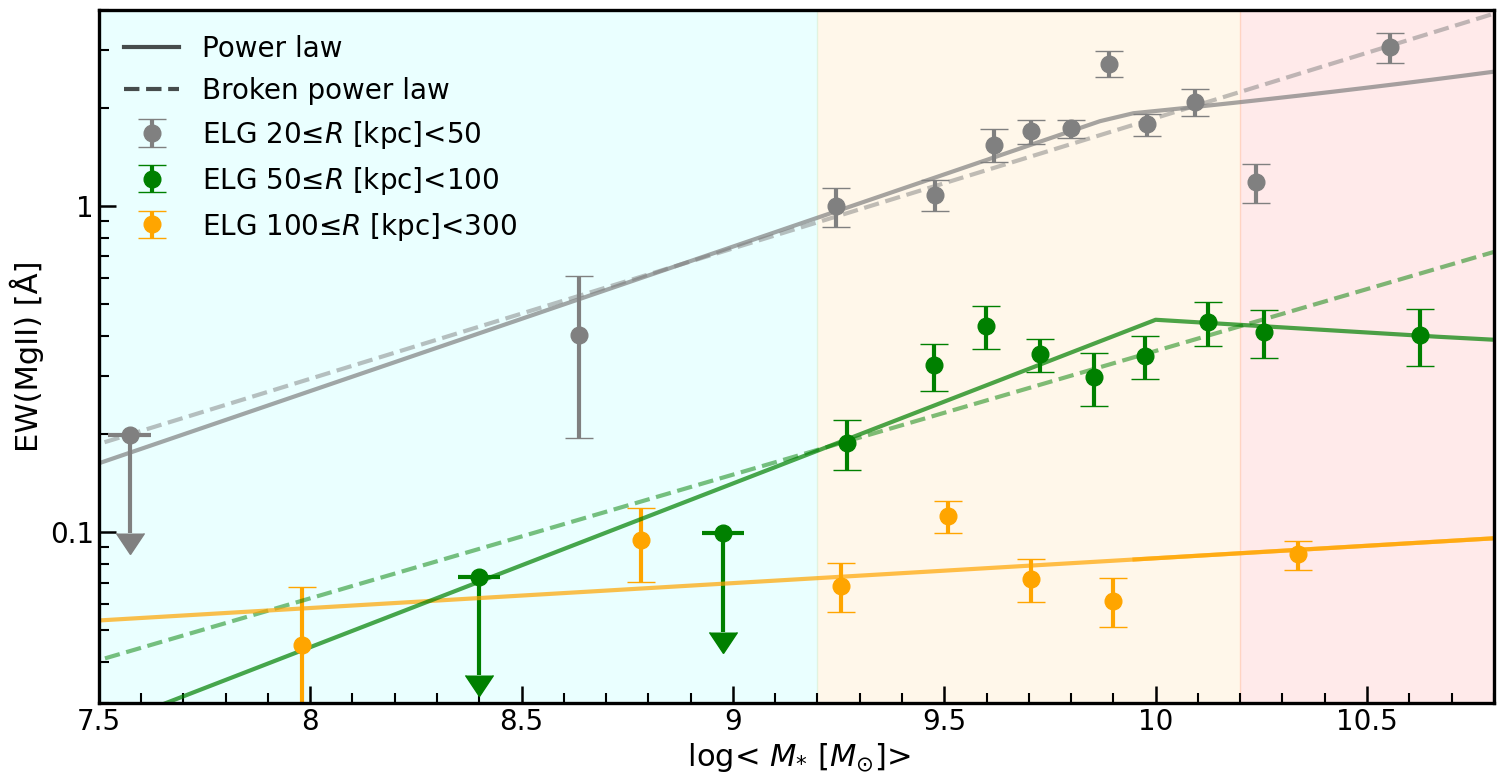}
    \caption{The \rm{EW(Mg\,\textsc{ii})} as a function of stellar mass measured in three impact parameter bins ($20\leq R\ [\rm{kpc}]<50$ , $50\leq R\ [\rm{kpc}]<100$, $100\leq R\ [\rm{kpc}]<300$) for ELGs. The fitted line is the result of a broken power law fit applied to the data points. For the grey and green fitted lines, we both identify a break point near $10^{10} M_{\odot}$, while the orange line does not have a break point. The expressions for the linear portions on the left side of the grey and green lines, as well as the orange line, are as follows: $\log (\rm{EW [\text{\AA}]} ) = 0.44\log(M_{*} [M_{\odot}]) - 4.11$ (grey), $\log (\rm{EW [\text{\AA}]} ) = 0.50\log(M_{*} [M_{\odot}]) - 5.36$ (green) and $\log (\rm{EW [\text{\AA}]} ) = 0.08\log(M [M_{\odot}]) - 1.84$ (orange).} The background colors in the figure indicate the three mass bins in Figure \ref{fig:elgew}.
    \label{fig:elgewm}
\end{figure*}


Stellar mass, as one of the most important parameters of galaxies, are expected to correlate with the properties of the CGM of host galaxies.  The existence of SFMS indicates that more massive galaxies should accrete cool gas at a higher rate to sustain the stronger star formation activities  \citep{Lilly-13, Stewart-17, Hafen-22, Wang-19, Wang-23}. The stronger star formation activities may cause stronger gas outflow \citep{Nelson-19, Peroux-20, weng24}, further imprinting on the properties of CGM.   


To understand the variation of Mg\,\textsc{ii} absorption with the $M_{*}$ of galaxies, we show the Mg\,\textsc{ii} absorption strength as a function of the impact parameter for ELGs in Figure \ref{fig:elgew} with separating galaxies into three mass intervals. These three stellar mass bins are labeled in Figure \ref{fig:elgew}, with the average stellar masses being $10^{8.83}M_{\odot}$ (blue), $10^{9.76}M_{\odot}$ (orange) and $10^{10.51}M_{\odot}$ (red). For comparison, we show the results of \cite{lan18} in gray shaded regions, which are based on the ELGs of eBOSS survey \citep{dawson16,myers15,prakash16,raichoor17}. 
In  Figure \ref{fig:elgew}  and throughout this work,  the data points with errors represent the good measurements of EW(Mg\,\textsc{ii}),  while the data points with downwards arrows represent the upper limit of EW(Mg\,\textsc{ii}).

It can be seen that the Mg\,\textsc{ii} absorption strength around ELGs decreases monotonically with $R$. For galaxies of intermediate and high masses, the relation follows a broken power law with $\rm{EW(Mg\,\textsc{ii})} \propto R^{-0.6}$ at impact parameter greater than 100 kpc, and  $\rm{EW(Mg\,\textsc{ii})} \propto R^{-1.7}$ at  $R$ less than 100  kpc. This flattening beyond 100 kpc is a real signal, as most of the data points show detections at or above the 3-sigma level.  For lowest stellar mass bin,  ELGs show a similar relation at small impact parameter,  although  the Mg\,\textsc{ii} absorption features at radius between 60-100 kpc are within the uncertainties of stacked spectra.   By comparing our results to the work of \cite{lan18},   we find general agreement in the observed trends of $\rm{EW(Mg\,\textsc{ii})}$ decreasing with impact parameter.  Since eBOSS survey is shallower than DESI survey,  the ELGs in eBOSS are biased to massive galaxies ($\sim 3 \times 10^{10}\thinspace \rm{M_{\odot}}$) with respect to DESI ELGs.  This is also the reason why the result of \cite{lan18} is in best agreement with our result for the galaxies in highest stellar mass bins.  

We emphasize that for both DESI and eBOSS ELGs,  the EW(Mg\,\textsc{ii}) as a function $R$ exhibit a shape of broken power law with the broken radius of 100-200 kpc. 
Speculating that this effect might be related to the size of host dark matter halos, we then estimate the average virial radius with $R_{\text{vir}} = 200 \, \text{kpc} \left({M_*}/{10^{10.3} \, {\rm M}_{\odot}} \right)^{1/5}$, provided by \cite{zhu13ca}. 
 

We indicate the virial radius of these galaxies with dashed lines in Figure \ref{fig:elgew} and find that the break radius of the power law approximately corresponds to the virial radius of the galaxies. 
This indicates that the distribution of cool gas inside and outside the dark matter halos exhibits different patterns. Inside halos, the density of cool gas decreases sharply with increasing radius, while outside halos, the distribution of cool gas is relatively sparse and relatively uniform and shows little dependence on stellar mass of host galaxies. It is also possible that the Mg\,\textsc{ii} absorption at $R>200$ kpc is related to galaxy groups \citep{Dutta-21, Zou-24b}, while central galaxies usually dominate the galaxy population at the mass range we considered \citep[e.g.][]{Wang-18}. Further investigations are needed in the future with larger samples to figure out the reason of this broken power law of EW(Mg\,\textsc{ii}).




A significant difference between this work and \cite{lan18} is that the deep DESI survey enables us to push the investigation of cool CGM to galaxies of lower stellar mass. We specifically show the EW(Mg\,\textsc{ii}) of ELGs as a function of stellar mass at three given intervals of impact parameter, in Figure \ref{fig:elgewm}. As we can see, within 100 kpc, there is a noticeable break in the power law at \(10^{10} M_{\odot}\), while beyond this range, no break point appears. So we perform a broken power-law fit to the data points within 100 kpc. The reduced chi-squared (\( \chi_\nu^2 \)) values are 0.017 (grey) and 0.015 (green), which are smaller than those of the single power-law fit, 0.023 (grey) and 0.019 (green), indicating that the broken power-law fit is more appropriate. The equivalent width of Mg\,\textsc{ii} absorption strongly depends on the mass of star-forming galaxies with $\rm{EW(Mg\,\textsc{ii})} \propto M_{*}^{0.5}$ for galaxies of $\log M_{*}/M_{\odot} < 10$ and $R< 100$ kpc, but becomes independent of mass for galaxies above this mass threshold. 
The flatten of EW(Mg\,\textsc{ii}) at high mass end is interesting and we will further study the possible origin of it in Section \ref{subsec:angledependence}. 
At impact factor greater than 100 kpc, the  \rm{EW(Mg\,\textsc{ii})}  appears to be nearly independent of stellar mass across the full range of mass we considered.  This supports the idea  that the Mg\,\textsc{ii} absorption beyond 100 kpc may not be closely related to the host dark matter halos.   

Using individual Mg\,\textsc{ii} absorbers,  \cite{anand21}  found that the ELGs do not exhibit an increase in  Mg\,\textsc{ii} covering fraction  with stellar mass at small scales and even show a slight decreasing trend, while the $\rm{EW(Mg\,\textsc{ii})}$ for
ELGs does not seem to depend on stellar mass (or may slightly
increase with it). They tend to conclude that Mg II absorption remains constant or changes little with stellar mass. Given that the median stellar mass for the their three bins are approximately $10^{9.9} M_{\odot}$, $10^{10.3} M_{\odot}$, and $10^{10.7} M_{\odot}$, and all fall near or on the flat section to the right of the turnover point in Figure \ref{fig:elgewm}, their finding that $\rm{EW(Mg\,\textsc{ii})}$ shows little variation with stellar mass is actually consistent with our results.

\begin{figure}[htbp]
    \centering
    \includegraphics[width=1
\columnwidth]{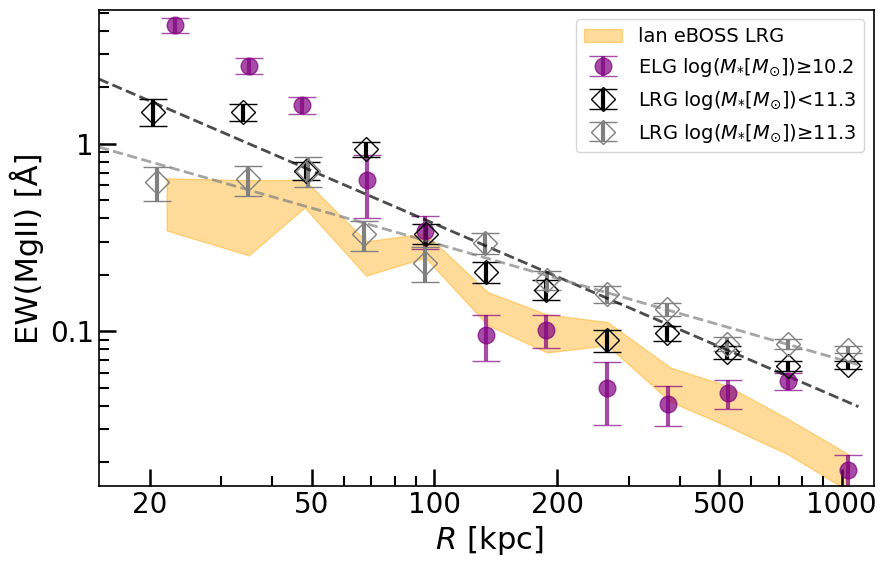}
    \caption{The \rm{EW(Mg\,\textsc{ii})} as a function of impact parameter for LRGs of different stellar masses. The average stellar masses and the fitting line for the two subsamples are: $10^{11.16}{\rm M}_{\odot}$, $\log (\rm{EW [\text{\AA}]}) = -0.94\log(R [\rm{kpc}]) + 1.45$  (black); $10^{11.46}{\rm M}_{\odot}$, $\log (\rm{EW [\text{\AA}]}) = -0.62\log(R [\rm{kpc}]) + 0.72$ (grey). The shaded orange area shows the result obtained by \cite{lan18}. The purple dot represents the result of ELGs in the highest stellar mass, as shown by the red point in Figure \ref{fig:elgew}.}
    \label{fig:lrgew}
\end{figure}

For completeness, we also study the  EW(Mg\,\textsc{ii}) of LRGs as a function impact factor, shown in Figure  \ref{fig:lrgew}.  
\cite{huang16} proposed that the presence of Mg\,\textsc{ii} absorbers surrounding LRGs can be accounted for by a mix of cool clouds that form within thermally unstable LRGs halos and the accretion of satellite material through filaments.  Therefore, the differences in the CGM properties from ELGs are expected.  

We separate the LRGs into two stellar mass bins with the averaged values being  $10^{11.16}M_{\odot}$ (black) and $10^{11.46}M_{\odot}$ (grey).  
For comparison, we show the result of ELGs of highest stellar mass bin (in Figure \ref{fig:elgew}) in purple data points.   Overall, the EW(Mg\,\textsc{ii}) tends to decrease with increasing $R$ for LRGs as well, following a power law, $R^{-0.94}$ (black) or $R^{-0.62}$ (grey).
In contrast to ELGs, the Mg\,\textsc{ii} absorption of LRGs is weaker at impact parameter less than 70 kpc, while becomes stronger at impact parameter greater than 100 kpc.  We note that the comparison here should be with caution, due to the different stellar mass ranges.  

Comparing with the result of \cite{lan18}, the DESI LRGs appear to show stronger Mg\,\textsc{ii} absorption than eBOSS LRGs across the full range of impact parameter we considered.   This is not due to the methodologies,  since we have reproduced the results of \cite{lan18} with eBOSS dataset as shown in Appendix \ref{sec:appenB}.   Therefore, the difference with \cite{lan18} in Figure \ref{fig:lrgew}  is likely from the different properties of LRG samples, such as stellar mass, redshift and etc.

\subsection{Evolution with redshift}\label{subsec:redshift}

\begin{figure*}[htbp]
\centering
    \includegraphics[width=0.9\columnwidth]{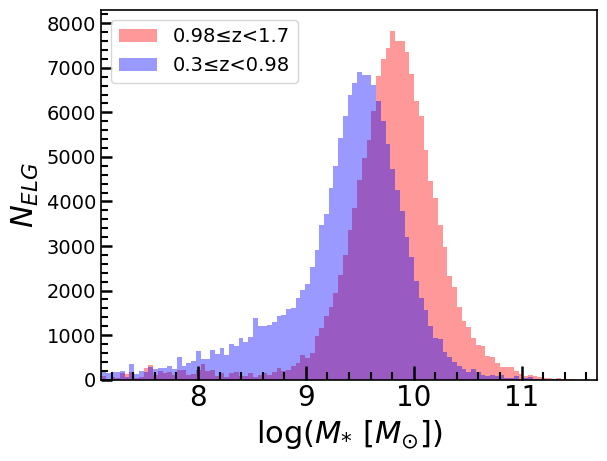}
    \includegraphics[width=1.04\columnwidth]{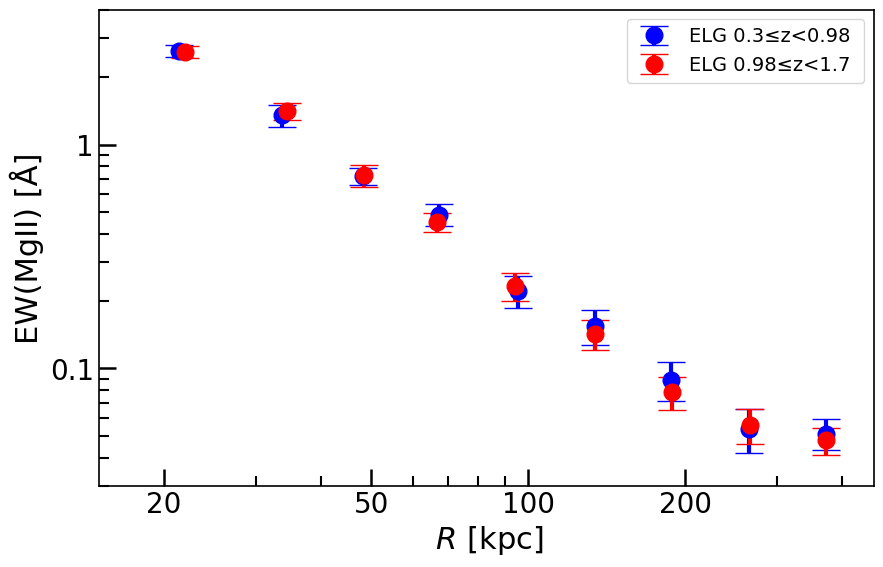}
    \caption{Left panel: The distribution of ELGs of two different redshift bins, as indicated in the figure. 
    We here only select galaxies in the overlapped regions with completely identical mass distributions. The average stellar masses for these two subsamples are $10^{9.80}M_{\odot}$. Right panel: The \rm{EW(Mg\,\textsc{ii})} as a function of impact parameter for ELGs in the two redshift bins.  }
    \label{fig:elgewz}
\end{figure*}

Whether the Mg\,\textsc{ii} absorption in CGM evolve with redshift or not is still under debate. According to \cite{matejek12}, the general structure of Mg\,\textsc{ii}-bearing halos was put into place early in the process of galaxy assembly and have evolved fairly little. Recently by using individual Mg\,\textsc{ii} absorbers,  \cite{anand21} found that a weak evolution in $\rm{EW(Mg\,\textsc{ii})}$ with galaxy redshift.  By separating individual absorbers into strong and weak populations ($W_{\rm \lambda 2796}>1\AA$ or not),   \cite{lan20} found that the  cover fraction for weak absorbers is consistent with being no redshift evolution, while strong absorbers around both star-forming and passive galaxies evolves significantly with redshift.  In this subsection, we for the first time examine the redshift dependence of  $\rm{EW(Mg\,\textsc{ii})}$ based on the stacked spectra, which is likely an unbiased tracer of CGM with respect to individual Mg\,\textsc{ii} absorbers.  

To do this,  we divide the ELGs into two subsamples of nearly equal number of galaxies with two redshift intervals, which are $0.3<z<0.98$ and $0.98<z<1.7$, respectively.   The stellar mass distribution of these two subsamples are shown in the left panel of Figure \ref{fig:elgewz}.  We then control the stellar mass distribution of two subsamples, by only selecting the overlapped regions in the mass distribution of Figure \ref{fig:elgewz}. 
In the overlapped regions, the average stellar mass is $10^{9.80}M_{\odot}$.  We then show the $\rm{EW(Mg\,\textsc{ii})}$ as a function impact factor for the two ELG subsamples of different redshifts, in Figure \ref{fig:elgewz}.  As can be seen, after eliminating the effect of mass, we find that the Mg\,\textsc{ii} absorption show no evolutionary trend with redshift, at least for the redshift range we considered here. However, \cite{wu2024} have used DESI-Y1 data to investigate the evolution of Mg\,\textsc{ii} absorbers with redshift, and found stronger absorption in the higher redshift bin in their stacking detection, especially for $R < 1$ Mpc. Considering that the stellar mass of ELGs in DESI increases with redshift (as shown in Figure \ref{fig:SFR}), this evolutionary trend may actually be driven by the difference of $M_*$, as the stellar mass distribution of samples within their same mass bin at different redshifts is not identical. And the median stellar mass in their higher redshift bin is clearly larger than that in lower redshift bin. 



\subsection{Dependence on SFR and $R_{\rm e}$ for ELG }\label{subsec:SFRandRe}

\begin{figure*}[htbp]
    \quad\quad
    \includegraphics[width=0.849\columnwidth]{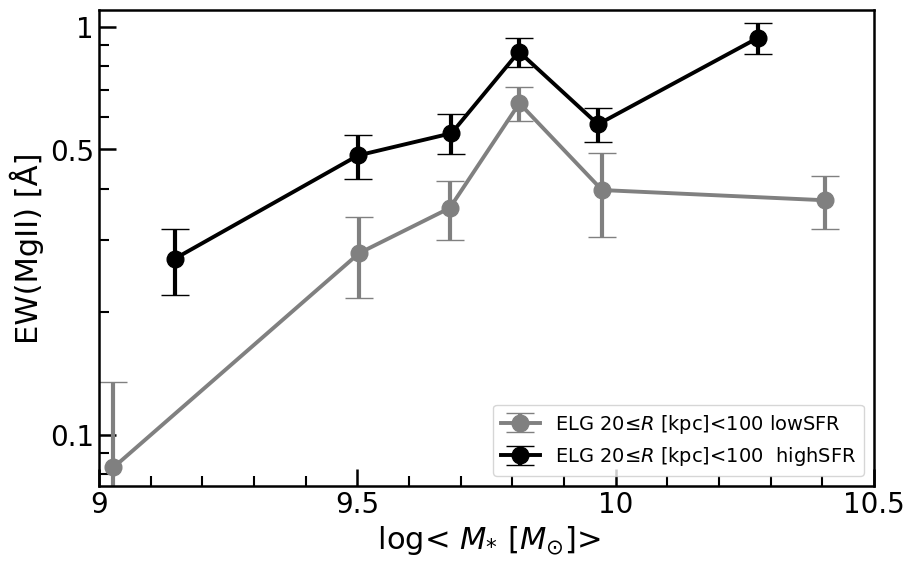}
    {\hspace{1.17cm}
        \includegraphics[width=0.849\columnwidth]{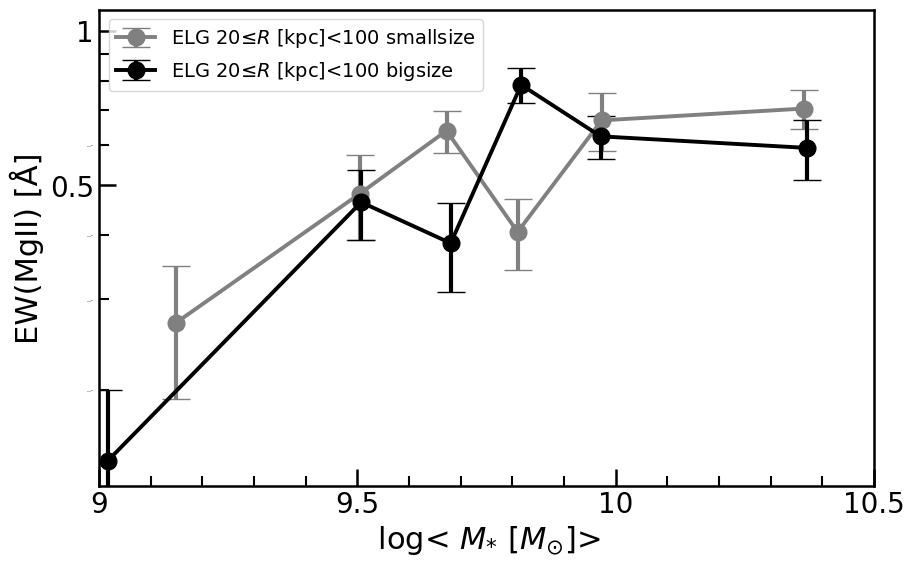}}
    \par 
    \quad\quad
    \includegraphics[width=1\columnwidth]{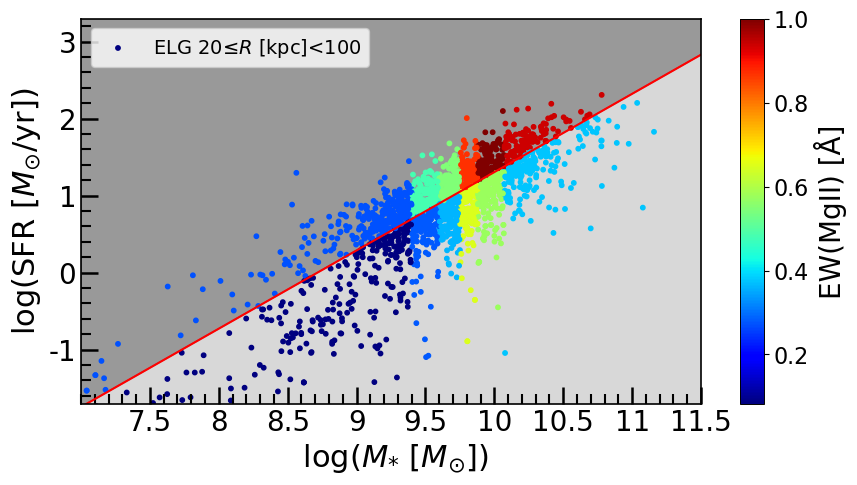}
    \includegraphics[width=1\columnwidth]{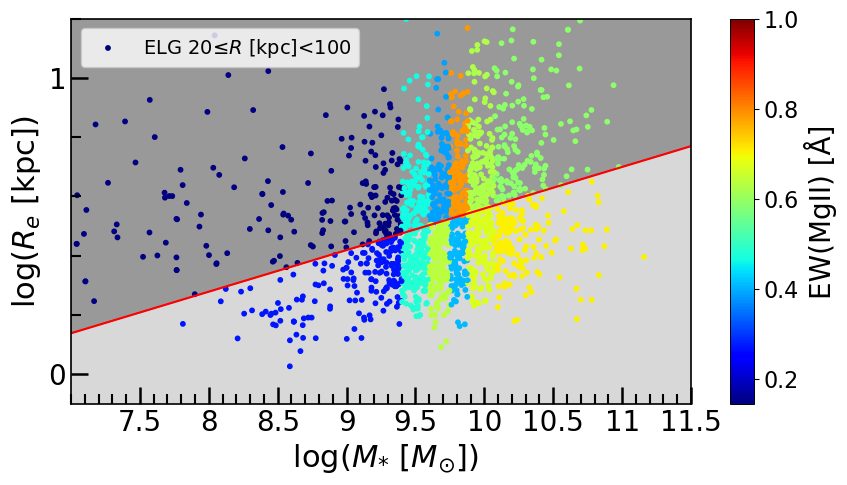}
    \caption{Top left: $\rm{EW(Mg\,\textsc{ii})}\ (20 < R [kpc]<100)$ for the two subsamples with SFR above (black) and below (grey) the fitted SFMS.  
    Top Right: $\rm{EW (Mg\,\textsc{ii})}\ (20<R [kpc]<100)$ for the two subsamples with $Re$ above (black) and below (gray) the fitted mass-size relation.  
    Bottom left: The sample definition of the top-left panel with the color-coding of $\rm{EW(Mg\,\textsc{ii})} \ (20<R [kpc]<100)$. 
    Bottom Right: The sample definition of the top-right panel with the color-coding of $\rm{EW(Mg\,\textsc{ii})} \ (20<R [kpc]<100)$.  
    }
    \label{fig:ELGfr1}
\end{figure*}

\begin{figure}[htbp]
    \includegraphics[width=0.9\columnwidth]{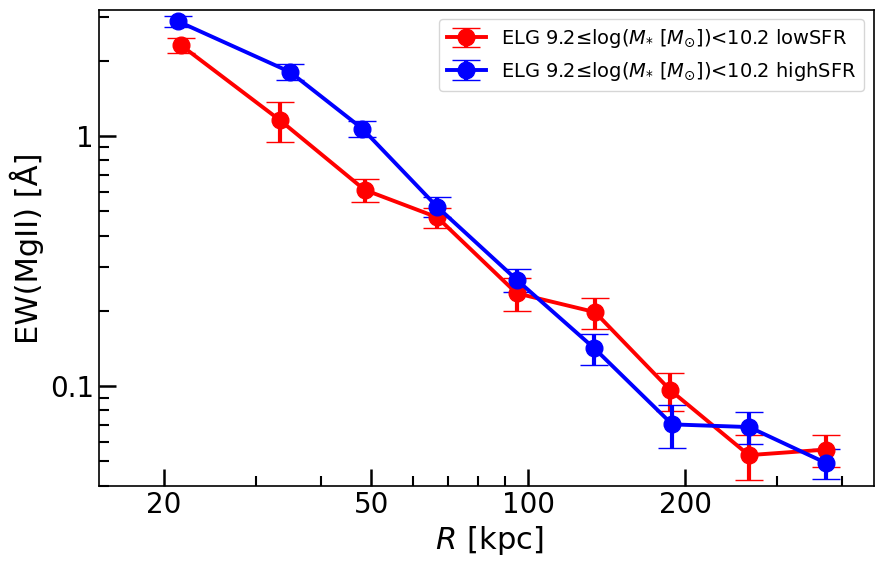}
    \par
    \includegraphics[width=0.9\columnwidth]{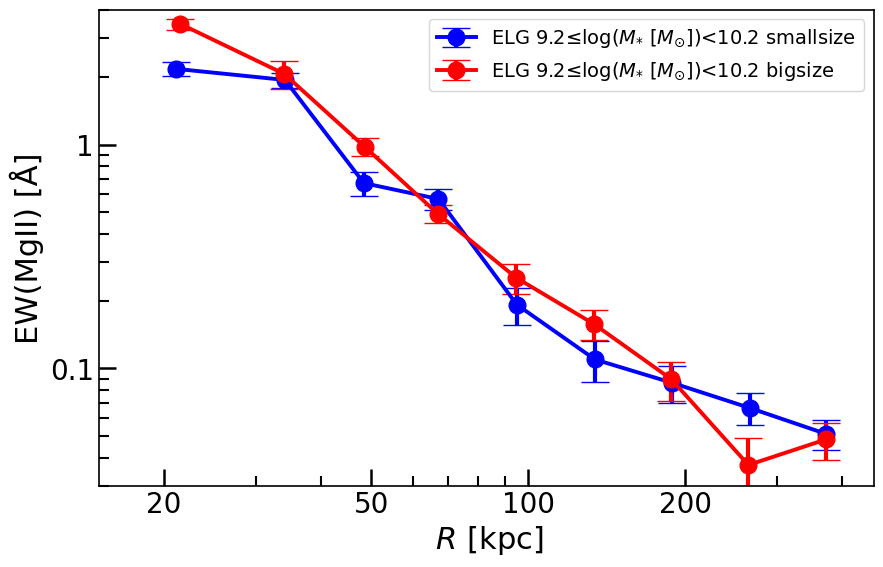}
    \caption{The $\rm{EW(Mg\,\textsc{ii})}$ as a function of impact parameter for samples with different SFR (top panel) and $R_{\rm e}$ (bottom panel).  The ELGs are separated according to the fitted SFMS and mass-size relation as shown in Figure \ref{fig:ELGfr1}.  Here we only use the ELGs with  $9.2 \leq \log(M_{*} [M_{\odot}])<10.2$.  }
    \label{fig:ELGfr2}
\end{figure}

Apart from $M_{*}$, SFR is also a key factor that influences the properties of Mg\,\textsc{ii} absorption lines \citep[e.g.][]{noterdaeme2010,kacprzak12,rubin18,huang2021,Zou-24a}, which is directly linked to the gas outflow associated with star formation. 
Indeed, as shown in Figure \ref{fig:elgew} and \ref{fig:lrgew}, we have shown that there are significant differences in the Mg\,\textsc{ii} absorption properties between star-forming galaxies (ELGs) and quenched galaxies (LRGs).  In consistent with this, quasar spectra with stronger absorption systems are more closely linked with bluer, star-forming galaxies \citep{zibetti07}. 

To provide a clear perspective on the influence of SFR on the behavior of Mg\,\textsc{ii} absorbers, we divided ELGs into two groups according to the fitted SFMS, as shown in the red line of the bottom-left panel in Figure \ref{fig:ELGfr1}.  The dividing line is obtained from fitting the median SFR values at different $M_{*}$ bins using the method of least squares. For each group, we further divide the ELGs into a set of stellar mass bins, ensuring that the number of spectra available for stacking is nearly equal in each bin, and is still enough to obtain a reliable signal in spectral stacking.


In the bottom-left panel of Figure \ref{fig:ELGfr1}, the colors indicate the average $\rm{EW (Mg\,\textsc{ii})}(20<R [kpc]<100)$ for galaxies that located on the different positions of SFMS.  As shown, overall, $\rm{EW (Mg\,\textsc{ii})}$ increase with stellar mass, and at given stellar mass, the $\rm{EW (Mg\,\textsc{ii})}$ is higher for galaxies of higher SFR.  To intuitively demonstrate the differences in $\rm{EW(Mg\,\textsc{ii})}$ between this two groups, we provide a concise visual representation in the top-left panel of Figure \ref{fig:ELGfr1}.

The positive correlation between Mg\,\textsc{ii} absorption strength and SFR is consistent with the previous findings of \cite{lan18} and \cite{anand21}. However, due to the limitations of sample size and the range of stellar mass of galaxies, they did not explore the impact of SFR with controlling $M_{*}$. As shown on the $M_*$-SFR diagram, ELGs with larger SFR tend to have larger stellar mass, making it difficult to separate the influences of SFR and $M_{*}$ as individual parameters. Thanks to the larger and deeper DESI survey, 
we are able to rigorously investigate the influence of SFR on Mg\,\textsc{ii} absorption within several narrow mass bins.  This pronounced absorption in ELGs of higher SFR reflect their more active gas outflow and richer gas supply, which is consistent with the predictions of simulations \citep[e.g.][]{Muratov-15, Nelson-19, Wang-23}.

We also examine whether the properties of Mg\,\textsc{ii} absorption in CGM of ELGs depend on the size of host galaxies or not. Similar to above approach, we separate ELGs into a set of stellar mass bins with larger and smaller $R_{\rm e}$, shown in the top/bottom-right panel of Figure \ref{fig:ELGfr1}.  As can be seen, overall we do not find clear evidence that the ELGs of different sizes show significantly different Mg\,\textsc{ii} absorption features.  



Apart from the disparities in absorption between ELGs with high and low SFRs, we are also interested in where these distinctions may persist in CGM, as this essentially defines the scope of the influence of star-forming activity. In Figure \ref{fig:ELGfr2}, we explore the $\rm{EW(Mg\,\textsc{ii})}$ as a function of impact parameter for ELGs of higher SFR and lower SFR, as classified in the bottom-left panel of Figure \ref{fig:ELGfr1}.  Considering the signal-noise-ratio of the data points, we only consider the ELGs that located in the stellar mass range of $10^{9.20}-10^{10.20}M_{\odot}$, and do not further separate ELGs into smaller mass bins.  

We find that the influence of SFR on $\rm{EW(Mg\,\textsc{ii})}$ is limited, typically within a range of approximately 50 kpc. Beyond this range, the absorption strengths start to converge for the two subsamples in the top panel of Figure \ref{fig:ELGfr2}.   Moreover, this 50 kpc range is in agreement with the previous findings \citep{heckman00, grimes05,rubin11,martin05}, suggesting that galactic outflows have velocities comparable to the escape velocity only at distances roughly 5–50 kpc from the starburst. 
This implies some correlation between SFR and outflows, which we will further explore in Section \ref{subsec:angledependence} by jointly considering the distribution of Mg\,\textsc{ii} absorption on azimuthal angle.

We also show the $\rm{EW(Mg\,\textsc{ii})}$ as a function of impact parameter for ELGs of different sizes in the bottom panel of Figure  \ref{fig:ELGfr2}.  Again, there is no clear evidence that the strength of Mg\,\textsc{ii} absorption significantly depends on the size of host galaxies. 
\subsection{Dependence on azimuthal angle}\label{subsec:angledependence}


The azimuthal angle ($\phi$) of the absorber, defined as the angle between the line connecting the absorber to the galaxy center and the major-axis of host galaxy, is a critical parameter for understanding the overall structure and dynamics of the gas circulation of galaxies.  
The work of \cite{bouche13} indicated that gas accretion along the major axis plays a significant role in galaxy growth because the estimated accretion rate is comparable to the SFR.  \cite{bordoloi14b} found significant variations in the distribution of cold gas within 40 kpc, depending on the azimuthal angle.  
Meanwhile, when gas is pushed out from a galaxy due to stellar feedback, it tends to align more with the galaxy's minor axis \citep[e.g.][]{Nelson-19}.   Consistent with this, the strongest Mg\,\textsc{ii} absorption occurs in bipolar regions that align with the poles of the disks \citep{kacprzak11,bordoloi11}, which could be evidence of gas outflowing along the minor axis of the galaxy.  Based on MEGAFLOW survey, \cite{schroetter19} found the strong bi-modal
distribution of Mg\,\textsc{ii} absorbers in the azimuthal angle, suggesting a scenario in which outflows are bi-conical and co-exist with a coplanar gaseous structure extending at least up to 60 kpc.  


\begin{figure}[htbp]
    \centering
    \includegraphics[width=1\columnwidth]{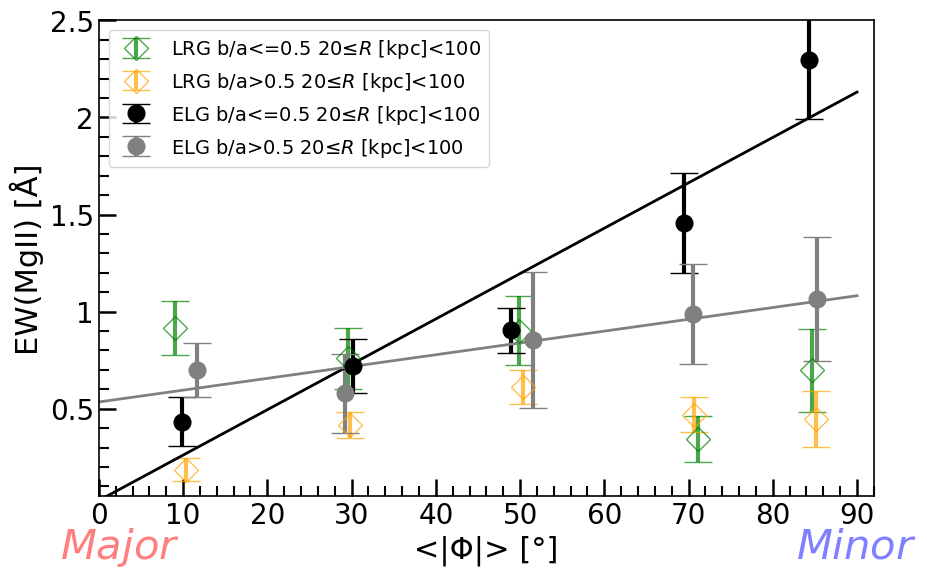}
    \caption{ The $\rm{EW(Mg\,\textsc{ii})} \ (20<R [kpc]<100)$ as a function of azimuthal angle for samples of different types, as indicated in the top-left corner.  In presenting the results, we separate ELGs (and LRGs) into two subsamples according to their minor-to-major axis ratio, $b/a>0.5$ or $b/a<0.5$. 
    The fitted lines for ELGs are: $\rm{EW [\text{\AA}]}=0.023<|\Phi|> $[°]$ +0.03$ (black) and $\rm{EW [\text{\AA}]}=0.006<|\Phi|> $[°]$ +0.53$ (grey).}
    \label{fig:ba}
\end{figure}

To further explore the anisotropic distribution of cool gas, we show the $\rm{EW(Mg\,\textsc{ii})}$ as a function of azimuthal angle in Figure \ref{fig:ba}, by dividing ELGs into two subsamples with $b/a>0.5$ (close to face-on) and $b/a<0.5$ (close to edge-on). 
As shown, for ELGs the strength of Mg\,\textsc{ii} absorption is stronger towards the minor axis of host galaxies.  Further more, this anisotropic distribution is more pronounced in ELGs with smaller $b/a$ with respect to those of larger $b/a$ as expected.   
This trend, however, has not been observed in LRGs (shown in green and orange), indicating that the dependence on azimuthal angle is originally from the star formation status of galaxies, such as galactic gas outflow as proposed in previous works \citep[e.g.][]{kacprzak11, bordoloi11, Nelson-19, schroetter19}.It is worth noting that the uncertainty of the position angle for weakly inclined galaxies (with larger $b/a$) will be larger compared to highly inclined galaxies (with smaller $b/a$), which may partially contribute to the isotropic distribution of absorption for galaxies with $b/a>0.5$. 


\begin{figure}[htbp]
    \centering
    \includegraphics[width=1\columnwidth]{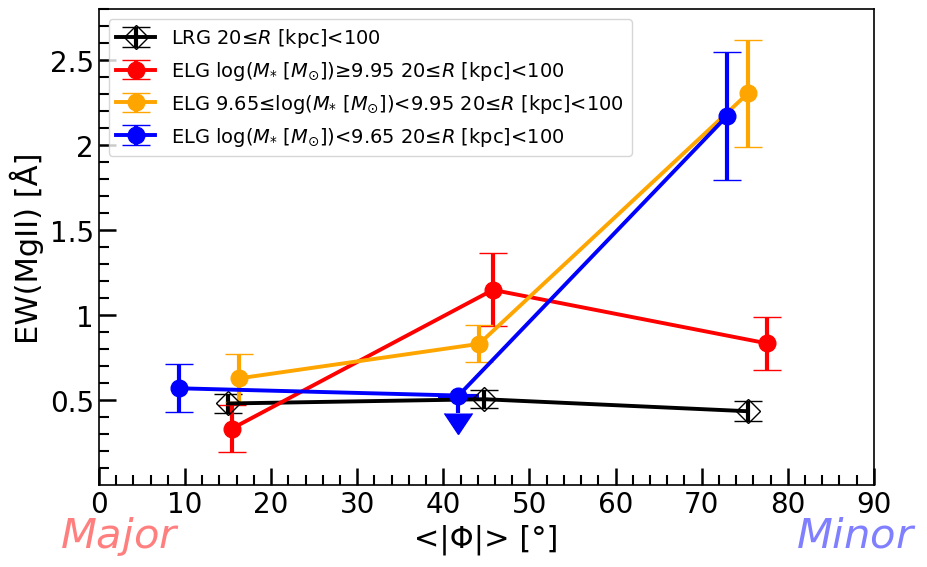}
    \caption{The angular distribution of $\rm{EW(Mg\,\textsc{ii})}\ (20 <R [kpc]<100)$ for LRGs and ELGs of three different stellar mass bins, as denoted at the top-left corner. 
    }
    \label{fig:mass}
\end{figure}

In Section \ref{subsec:RMdependence}, we find that the $\rm{EW(Mg\,\textsc{ii})}$ increases with stellar mass at low mass end, but flattens at the stellar mass above $\sim10^{10} {\rm M_{\odot}}$. Here, we try to explore the possible reason for the flattening of $\rm{EW(Mg\,\textsc{ii})}$ at high mass end.  
We therefore investigate the $\rm{EW(Mg\,\textsc{ii})}$  as a function azimuthal angle, by separating the ELGs into several stellar mass bins. As illustrated in Figure \ref{fig:mass}, there are three mass bins. The most massive bin is designed to facilitate the study of the angular distribution of $\rm{EW(Mg\,\textsc{ii})}$ for ELGs with masses greater than the break point ($10^{9.95}M_{\odot}$) in Figure \ref{fig:elgewm}. And the other two bins are selected to ensure reasonable signal strength with comparable S/N ratios at the low-mass end.  
As can be seen, there is an excess of Mg\,\textsc{ii} absorption close to the minor axis of ELGs with $M_*<10^{9.95} {\rm M_{\odot}}$,  while this is not seen for ELGs of stellar mass above this threshold.  This suppression of Mg\,\textsc{ii} absorption near the minor axis of massive ELGs likely explain the flattening of  $\rm{EW(Mg\,\textsc{ii})}$ at the high stellar mass end in Figure \ref{fig:elgewm}. 

The reason of the suppression may due to two things. One is the decreasing mass-loading factor with increasing stellar mass, since deeper gravitational potential would suppression the gas outflow at minor axis. The other is that the outflowing gas may not in neutral phase, which can be heated by the feedback of AGN with increasing stellar mass.   
More analyses are needed in the future with larger dataset to further figure this out.  



\begin{figure}[htbp]
    \includegraphics[width=0.95\columnwidth]{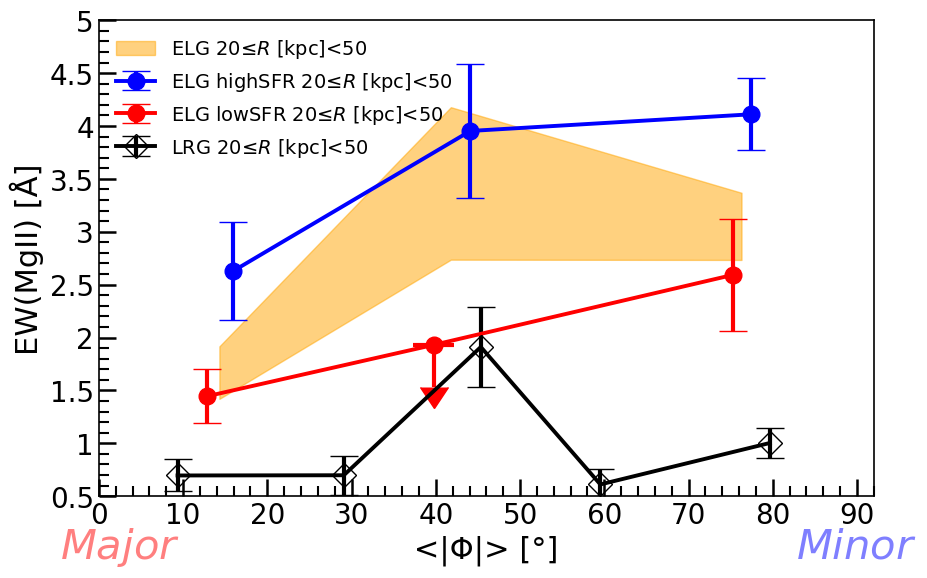}
    \par
    \includegraphics[width=0.95\columnwidth]{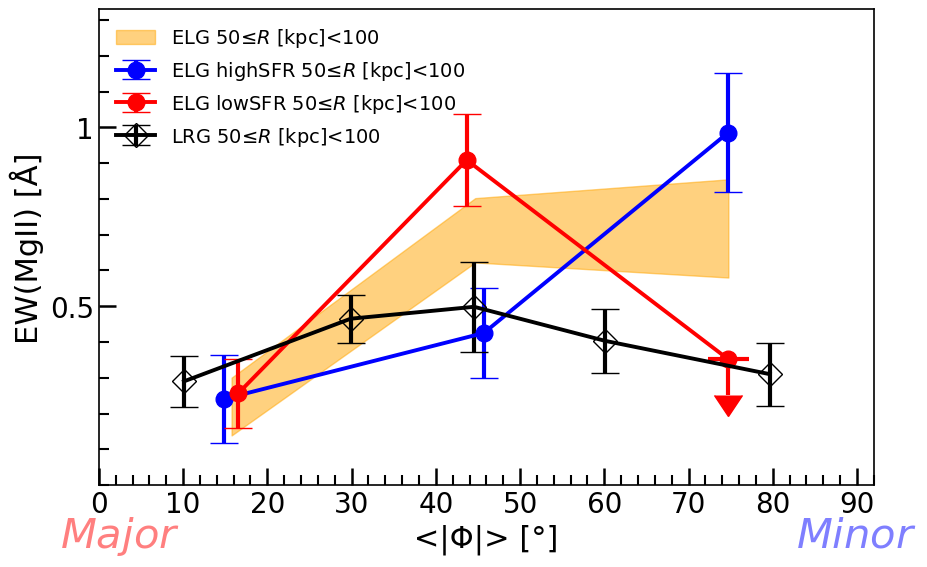}
    \caption{Angular distribution of $\rm{EW(Mg\,\textsc{ii})}$ measured at $20 <R\ [{\rm kpc}]<50$ (top) and at $50<R\ [{\rm kpc}]<100$ (bottom). In each panel, we show the results of ELGs (orange shaded region), and the two ELG subsamples with higher SFR (blue) and lower SFR (red).  For comparison, the result of LRGs is shown in black. 
    }
    \label{fig:angle}
\end{figure}

Additionally, we explore the azimuthal distribution of $\rm{EW(Mg\,\textsc{ii})}$ for galaxies of different SFRs by categorizing ELGs (shaded area) into two groups based on whether their SFRs are greater (or less) than the median values in each azimuthal angle bin.  The results are shown in Figure \ref{fig:angle}, from which we can read out the following interesting things.  
ELGs of higher SFRs show stronger Mg\,\textsc{ii} absorption at all spatial orientations, suggesting that they are more gas-rich in their gas reservoirs with respect to ELGs of lower SFR (and LRGs).   
ELGs of higher SFRs show excess Mg\,\textsc{ii} absorption nearly the minor axis even out to 50-100 kpc, confirming that higher SFR causes stronger galactic ouflow of cool gas.  
It is worth noting that LRGs (and ELGs of lower SFR at impact parameter of 50-100 kpc) appear to show an excess of Mg\,\textsc{ii} absorption at the azimuthal angle near 45 degree. This feature is potentially interesting, and the reason of it needs further investigation. 

Overall, we conclude that our result supports that the azimuthal dependence of cool CGM is primarily determined by the orientation of inflow and outflow of galaxies, as proposed by many previous results \citep{kacprzak12, bouche12, chisholm15, schroetter19}.  In this work we quantify this effect by looking at the median Mg\,\textsc{ii} absorption around galaxies with big sample, which provides important constraints in the feedback processes of  hydro-dynamic simulations.

\section{DISCUSSION} \label{sec:discussion}
\subsection{Comparison with individual absorbers}\label{subsec:coverfraction}

\begin{figure*}[htbp]
    \centering
    \includegraphics[width=2\columnwidth]{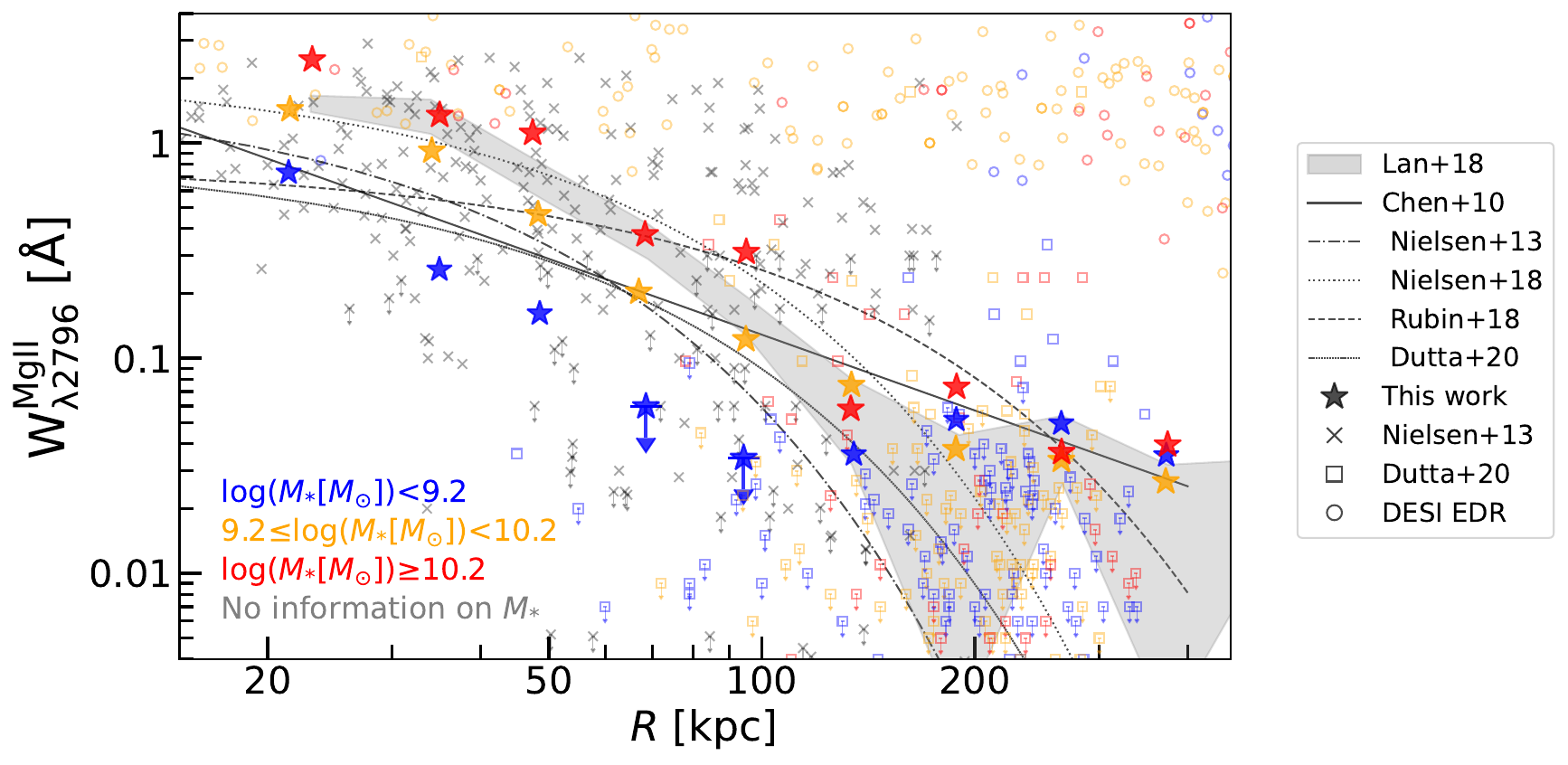}
    \caption{The $\rm{W}_{\lambda 2796}^{\text{Mg\,\textsc{ii}}}$ as a function of impact parameter for ELGs of different stellar mass bins. We distinguish these Mg,\textsc{ii} absorbers from different studies by shapes, while colors represent the $M_{*}$ of their host galaxies. Our stacked results are shown using stars, while the gray crosses represent data from \cite{nielsen13}, and the squares represent data from \cite{dutta20}. The circles represent the Mg\,\textsc{ii} absorbers that we have matched to the DESI's Mg\,\textsc{ii} catalog \citep{napolitano23} around  all ELGs from DESI EDR. All data points with downward arrows indicate that they are upper limits. 
   The gray lines of various styles are best-fitting relationships for individual Mg\,\textsc{ii} absorbers in the literature \citep{chen10absorb,nielsen13b,nielsen18,rubin18,dutta20}. 
   }
    \label{fig:elgfc}
\end{figure*}

In this section, we compare the $\rm{W}_{\lambda 2796}^{\text{Mg\,\textsc{ii}}}$ (the equivalent width for only the Mg\,\textsc{ii}\thinspace$\lambda 2796$ line) in our co-added spectra for ELGs with individual Mg\,\textsc{ii} absorbers associated with galaxies \citep{dutta20,nielsen13}. In the same way as in Section \ref{subsec:RMdependence}, we show the results by  separating ELGs into three different stellar mass bins, as labeled in Figure \ref{fig:elgfc}.  For comparison, the results of \cite{lan18} are also annotated with shaded areas. As shown, the $\rm{W}_{\lambda 2796}^{\text{Mg\,\textsc{ii}}}$ decreases with increasing impact parameter, and with decreasing stellar mass of host ELGs.  

Note again that the $\rm{W}_{\lambda 2796}^{\text{Mg\,\textsc{ii}}}$  with co-added spectra better represent the overall distribution of Mg\,\textsc{ii} absorption, which is primarily contributed by weaker Mg\,\textsc{ii} absorbers at large impact parameter.  The individual Mg\,\textsc{ii} absorbers may be biased to strong absorbers set by the detection limits of individual spectra. In Figure \ref{fig:elgfc}, we present the Mg\,\textsc{ii} absorbers from different studies with different shapes ($\star$: this work; $\times$:  \cite{nielsen13}; \scalebox{0.8}{$\square$}: \cite{dutta20}; \scalebox{1.5}{$\circ$}: DESI EDR), with color-coding of the stellar mass of their host galaxies (blue:\thinspace$M_{*} <10^{9.2}M_{\odot}$; orange:\thinspace$10^{9.2}M_{\odot}\leq M_{*} <10^{10.2}M_{\odot}$; and red:\thinspace$M_{*} >10^{10.2}M_{\odot}$). 
The circles (\scalebox{1.5}{$\circ$}) represent the Mg\,\textsc{ii} absorbers that we have matched to the DESI's {\tt \href{https://data.desi.lbl.gov/public/edr/vac/edr/mgii-absorber/v1.0/MgII-Absorbers-EDR.fits}{Mg\,\textsc{ii} catalog}} \citep{napolitano23} around all ELGs from DESI EDR. When conducting this matching, we utilize
 a $v_{\text{off}} = c \, (z_{\text{Mg II}} - z_{\text{galaxy}}) / (1 + z_{\text{galaxy}}) < 500$ km/s as the criterion to associate the absorbers with galaxies closely. Moreover, we display best-fitting relationships of these absorbers in the literature indicated by black lines of various styles \citep{chen10absorb,nielsen13b,nielsen18,rubin18,dutta20}. 

As shown in Figure \ref{fig:elgfc}, in cases where the $M_{*}$ of the host galaxies is approximately the same, our results are consistent with those of \cite{dutta20},  while the \(\rm{W}_{\lambda 2796}^{\text{Mg\,\textsc{ii}}}\) results from DESI's Mg\,\textsc{ii} catalog show a general offset toward higher values. This may be due to the lower signal-to-noise ratio of the DESI spectra, which results in only absorbers with higher absorption strengths being detectable. And this also reminds us that for studies of individual absorbers, the selection effects differ from survey to survey and source to source, so such comparisons should be made with caution.  However, the stacking technique provides quite consistent results, based on different surveys and different groups.  This suggests the superiority of the spectral stacking approach in studying the absorption of CGMs, although this is  more complicated than studying individual Mg\,\textsc{ii} absorbers.    


\subsection{Mg\,\textsc{ii} cover fraction and mass of circumgalactic neutral hydrogen}\label{subsec:MHI}

Mg\,\textsc{II} covering fraction ($f_{\rm c}$)  
is defined as the fraction of QSO sightlines in a given radial bin $R$ that have one or more absorbers satisfying a chosen $\rm{W}_{\lambda 2796}^{\text{Mg\,\textsc{II}}}$ threshold. According to \cite{prochaska14}, the average absorption around ELGs and LRGs is dominated by strong Mg\,\textsc{II} absorbers with \( W_{\lambda2796}^{\text{Mg\,\textsc{II}}} > 0.4 \, \text{Å} \), and with only a negligible contribution from weaker components. Therefore, in our co-added  spectra, we can calculate the Mg\,\textsc{II} absorption fraction ($>0.4\thinspace$\text{\AA}) by measuring the EW of the absorption line at 2796 Å in the Mg\,\textsc{II} doublet, following the relation \citep{lan18}:  
\begin{equation}
\langle W_{\lambda 2796}^{\text{Mg\,\textsc{ii}}} \rangle(r_p) \approx f_c(W_{\lambda 2796}^{\text{Mg\,\textsc{ii}}} > 0.4 \, \text{\AA}, r_p)
\times \hat{W}_{\lambda 2796}^{\text{Mg\,\textsc{ii}}},
\label{eq:fracMg}
\end{equation}
where $\langle W_{\lambda 2796}^{\text{Mg\,\textsc{ii}}} \rangle(r_p)$ represents the average profile we are examining, while $\hat{W}_{\lambda 2796}^{\text{Mg\,\textsc{ii}}}$ denotes the average of $\rm{W}_{\lambda 2796}^{\text{Mg\,\textsc{ii}}}$ values derived from individual absorbers along random QSO sightlines. And the $f_c(W_{\lambda 2796}^{\text{Mg\,\textsc{ii}}} > 0.4 \, \text{\AA}, r_p)$  here refers the covering fraction of absorbers with $W_{\lambda 2796}^{\text{Mg\,\textsc{ii}}} > 0.4 \, \text{\AA}$ within a narrow ring at a distance $r_{p}$ from the galaxy.
To calculate $\hat{W}_{\lambda 2796}^{\text{Mg\,\textsc{ii}}}$, we utilize the incidence rate, ${d^2N}/{dWdz}$, of individual Mg\,\textsc{ii} absorbers sourced from \cite{zhu13}.  For absorbers with $\rm{W}_{\lambda 2796}^{\text{Mg\,\textsc{ii}}} > 0.4$ \AA, the resulting $\hat{W}_{\lambda 2796}^{\text{Mg\,\textsc{ii}}}$ hovers around 1 \AA.

With the covering fraction obtained above, we can further estimate the amount of hydrogen mass around galaxies following the work of \cite{lan14}:
\begin{equation}
M_{\text{H\,\textsc{i}}}(r_{\text{p}} < R) \sim
2\pi \, m_{\text{H}} \, 
 \int_{10 kpc}^{R} \hat{N}_{\text{H\,\textsc{i}}} f_{c}( r_{\text{p}}) \, r_{\text{p}} \, dr_{\text{p}},
\label{eq:MH}
\end{equation}
where $\hat{N}_{\text{H\,\textsc{i}}}$ is the neutral hydrogen column density traced by the Mg\,\textsc{ii} absorbers and the $f_{c}( r_{\text{p}})$ is the same one ($f_c(W_{\lambda 2796}^{\text{Mg,\textsc{ii}}} > 0.4 \text{\AA}, r_p)$) that we obtained from Equation \ref{eq:fracMg}.
 Adopting the empirical relation between the rest equivalent width of Mg\,\textsc{ii} and $N_{\rm H\,\textsc{i}}$ derived by \cite{lan17}, we obtain $\hat{N}_{\text{H\,\textsc{i}}} \approx 3\times 10^{19}\rm{cm}^{-2}$ for $\hat{W}_{\lambda 2796}^{\text{Mg\,\textsc{ii}}} \sim 1 \, \text{\AA}$ at redshift 0.8. 

\begin{figure}[htbp]
    \includegraphics[width=0.9\columnwidth]{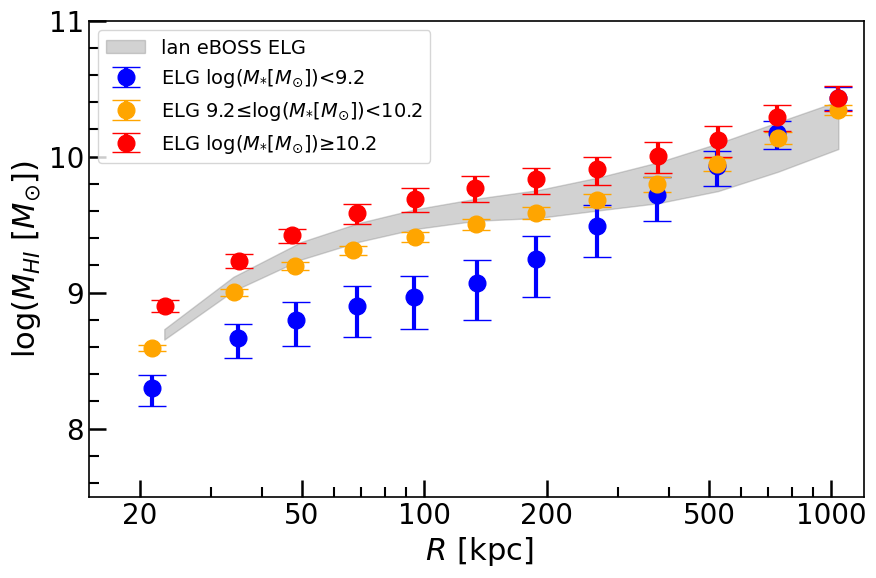}
    \includegraphics[width=0.9\columnwidth]{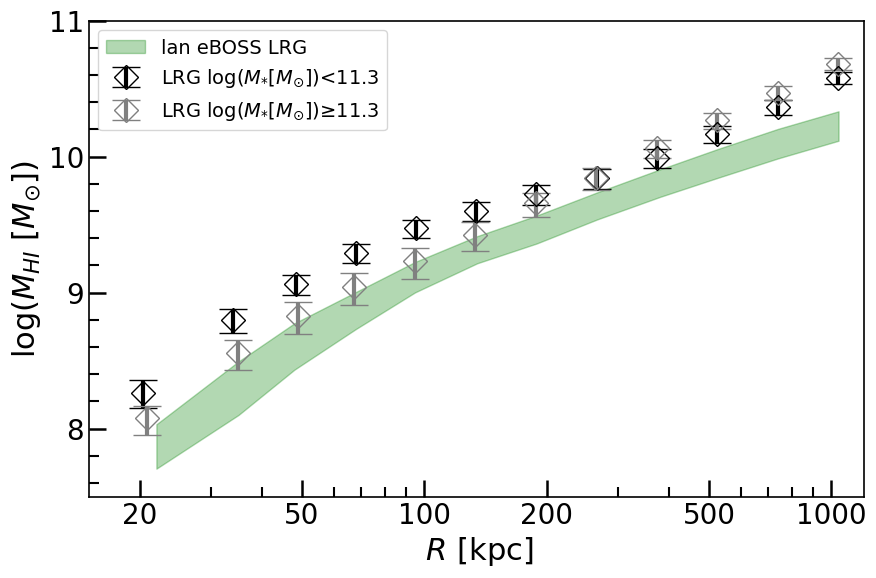}
    \caption{The cumulative $M_{\rm H\,\textsc{i}}$ calculated from $f_c(W_{\lambda 2796}^{\text{Mg,\textsc{ii}}} > 0.4 \text{\AA}, r_p)$  as a function of impact parameter for ELGs (top) and LRGs (bottom) of different stellar masses. The colors and symbols here are the same as those in Figure \ref{fig:elgew} and Figure \ref{fig:lrgew}.}
    \label{fig:HI}
\end{figure}

Our investigation reveals that within a distance of 200 kpc, the $M_{\rm H\,\textsc{i}}$ in the CGM demonstrates a positive correlation with the $M_{*}$ of the ELGs.   However, for LRGs, as shown in the bottom of Figure \ref{fig:HI}, this result is exactly the opposite, as more massive LRGs have less $M_{\rm H\,\textsc{i}}$ within this distance. This could be explained by \cite{wang2018sdss}, who suggested that quenching mechanisms become more prominent in higher-mass galaxies, which may lead to a reduction in $M_{\rm H\,\textsc{i}}$.  As depicted in the top of Figure \ref{fig:HI}, the $M_{\rm H\,\textsc{i}}$ within the lowest mass bin notably differs from the other two bins in $R < 200\rm{kpc}$. However, with increasing the impact parameter, the distinctions among the three bins rapidly diminish beyond 200 kpc. This observation suggests that at larger spatial scales, the $M_{\rm H\,\textsc{i}}$ of the CGM may predominantly be influenced by the surrounding environment rather than solely by the galaxies themselves. 

In fact, galaxy environment plays a crucial role in determining the large-scale structure of CGM. \cite{bordoloi11} found that Mg\,\textsc{ii} absorption around group galaxies is more extended than that for isolated galaxies, and interestingly, these effects can be satisfactorily modeled by a simple superposition of the absorption profiles of individual member galaxies. In light of this, it worth to further explore the CGM of galaxies with considering galaxy environment in the future, such as being centrals or satellites. 


\section{Summary} \label{sec:summary}
This work investigates the CGM properties of galaxies using Mg\,\textsc{ii} absorption lines from background QSOs through a spectral stacking technique. We utilize EDR of the largest on-going spectroscopy survey DESI, focusing on the CGM of ELGs.    
With the large amount of high-quality spectra, we are able to gain detailed insights into the distribution of Mg\,\textsc{ii} absorbers in the CGM of galaxies. The main findings are listed as follows: 

\begin{itemize}

\item

For ELGs, the \rm{EW(Mg\,\textsc{ii})} strongly correlates to the stellar mass following \rm{EW(Mg\,\textsc{ii})} $\propto M_{*}^{0.5}$ for galaxies of $\log M_{*}/M_{\odot} < 10$ and $R< 100$ kpc, but becomes independent with $M_*$ for galaxies above $\sim 10^{10}\ {\rm M_{\odot}}$ (Figure \ref{fig:elgewm}). 
The \rm{EW(Mg\,\textsc{ii})} as a function of impact parameter can be characterized as a broken power-law at the radius of 100-200 kpc (Figure \ref{fig:elgew}).  This radius can serve as a dividing line between the CGM associated with host galaxies and that associated with the surrounding environments.  Different with ELGs, LRGs do not show this broken radius in  \rm{EW(Mg\,\textsc{ii})} as a function of impact parameter.

\item
We examine the dependence of the \rm{EW(Mg\,\textsc{ii})} on redshift, and find that after controlling stellar mass, no clear evidence for significant evolution is seen in the redshift range we considered (Figure \ref{fig:elgewz}).

\item
We explore the dependence of Mg\,\textsc{ii} absorption on SFR and size of host ELGs. At given stellar mass, EW(Mg\,\textsc{ii}) is larger for ELGs of higher SFR at the impact parameter of less than 50 kpc, while shows little dependence on galaxy size (Figure \ref{fig:ELGfr1}). 


\item 
ELGs show stronger Mg\,\textsc{ii} absorption near the minor axis especially for strongly-inclined galaxies (Figure \ref{fig:ba}), while LRGs show little dependence of EW(Mg\,\textsc{ii}) on azimuthal angle. This indicates the connection with the gas outflow associated with star formation.   For ELGs with stellar mass above $\sim 10^{10}\thinspace {\rm M_{\odot}}$,  the Mg\,\textsc{ii} absorption at the minor-axis is largely suppressed, suggesting that the cool outflowing gas is suppressed or even disrupted.  This is likely related to the flattening of \rm{EW(Mg\,\textsc{ii})} for ELGs at the high mass end in Figure \ref{fig:elgewm}. 


\end{itemize}

Overall, our result indicates that the CGM distribution of ELGs within 100 kpc are strongly linked to the properties of host galaxies, likely in terms of gas accretion and bi-polar gas outflow.  However, both processes are directly related the stellar mass of host galaxies.  For instance, more massive star-forming galaxies tend to have higher inflow rate of cool gas, due to deeper gravitational potential. More massive star-forming galaxies also tend to have higher SFR, further leading to stronger gas outflow. However, the deeper gravitational potential will suppress this process. Therefore, the gas outflow at the minor-axis of galaxies is determined by these two competing effects.  


For impact parameter greater than 200 kpc, the CGM of ELGs appears to be more closely related to the surrounding environment out of the dark matter halos, rather than the host galaxies.   This is different to LRGs, where Mg\,\textsc{ii} absorption does not appear to be closely linked to the host galaxies, but to the host dark matter halos \citep{zhu14}, even at small impact parameter.  Here we only present a possible explanation to link these observational facts, while the up-coming larger datasets would help us to understand the distribution of cool CGM of galaxies in more details.  

\section*{Acknowledgements}
EW thanks the support of the National Science Foundation of China (Nos. 12473008) and the Start-up Fund of the University of Science and Technology of China (No. KY2030000200).
HYW is supported by the National Natural Science Foundation of China (Nos. 12192224) and CAS Project for Young Scientists in Basic Research, Grant No. YSBR-062. 
HZ acknowledges the supports from National Key R\&D Program of China (grant Nos. 2022YFA1602902, 2023YFA1607800, 2023YFA1607804) and the National Natural Science Foundation of China (NSFC; grant Nos. 12120101003 and 12373010).
SZ acknowledges support from the National Science Foundation of China (no. 12303011). 
YG receives funding from Scientific Research Fund of Dezhou University (3012304024), Shandong Provincial Natural Science Foundation (ZR2024QA212), and the National Natural Science Foundation of China (NSFC, Nos.12033004 and 12233005).
The authors gratefully acknowledge the support of Cyrus Chun Ying Tang Foundations.

\bibliography{main}
\bibliographystyle{aasjournal}
\appendix

\section{SFRs of DESI galaxies} \label{sec:appenD}

We calculate the SFRs of the galaxies using the luminosity of [O\,\textsc{ii}] emission line as described in \cite{calzetti00,schroetter19}, assuming an empirical relation of intrinsic dust extinction (see Equation \ref{eq:ebv}). We compare the SFMS of DESI ELGs obtained in the present work with the recent result obtained by \cite{speagle14} in Figure \ref{fig:D1}.  The solid lines represent the median SFRs of the DESI ELGs, separated into three redshift bins: $0.25\leq z<0.75$ (blue),  $0.75\leq z<1.25$ (orange), and $1.25\leq z<1.75$ (red). The dotted lines indicate the evolutionary trends of star-forming galaxies provided by \cite{koprowski24}.
As can be seen, the two are in broad agreement, indicating that the measurements of SFR in this work are reasonable. 

\renewcommand\thefigure{A}
\begin{figure}[htbp]
    \centering
    \includegraphics[width=1\columnwidth]{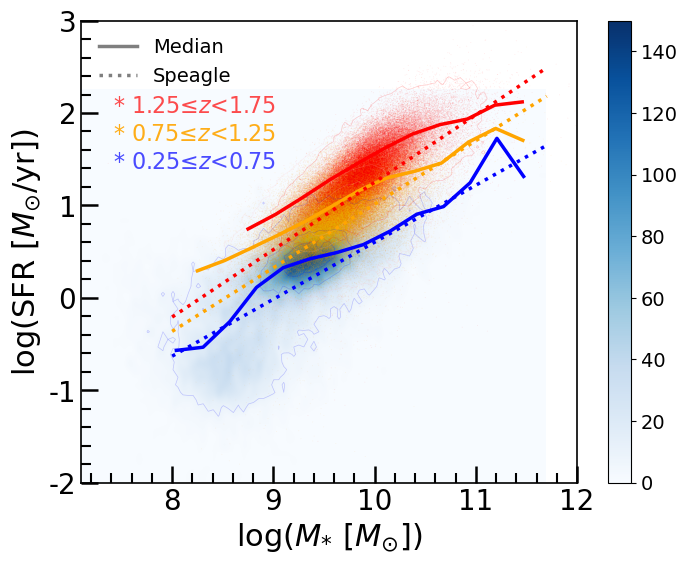}
    \caption{The star formation main sequence of DESI ELGs at different redshift intervals. The solid line represents the median SFR of the DESI ELGs, while the dotted line shows the evolutionary trend of star-forming galaxies given by \cite{speagle14}.}
    \label{fig:D1}
\end{figure}

\section{The distribution of impact parameter for QSO-galaxy pairs} \label{sec:appenA}

In Figure \ref{fig:A1}, we show the number distribution ($N_{pair}$) of the  QSO-galaxy pairs as a function of impact parameters. 
As shown, the number $N_{pair}$ linearly correlates with the impact parameter up to 1200 kpc. This is the nature prediction assuming that the distributions of background QSO have no correlation with the foreground galaxies.  
Therefore, the number of QSO contained within each annulus is proportional to the area of the annulus. This also confirms that the matching of QSO and galaxies are right.
\renewcommand\thefigure{B}
\begin{figure}[htbp]
    \centering
    \includegraphics[width=0.9\columnwidth]{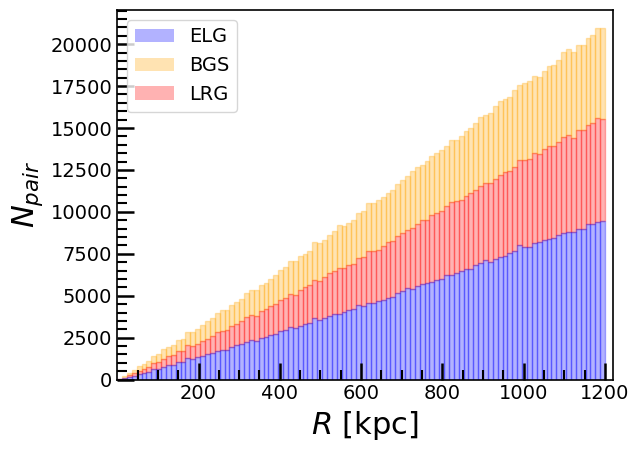}
    \caption{The distribution of the impact parameters for the QSO-ELG pairs (blue), QSO-LRG pairs (red), and QSO-BGS pairs (orange).}
    \label{fig:A1}
\end{figure}

\section{The process of handling a single spectrum}\label{sec:appenE}

Here we illustrate the detailed process of processing individual spectra in \ref{fig:spec}, starting from the original QSO spectrum and ending up with a normalized spectrum that is used for stacking.
In the top panel of Figure \ref{fig:spec}, we show a QSO spectrum in black, while the fitted spectrum in blue.  As can be seen, the fitted spectrum can very well characterize the overall features of the observed spectrum.  We also present the residuals between the original spectrum and the QSOfit (purple) and zoom in on some absorption lines (Fe\,\textsc{ii} \& Mg\,\textsc{ii}) and QSOfits at that location.
In the bottom panel of Figure \ref{fig:spec}, we show the normalized spectrum in gray, by dividing the original spectrum with the fitted spectrum.  
This step is crucial for removing the intrinsic features of QSO and isolating the absorption features. 
However, we still see some small-scale features in the normalized spectrum.  We then apply a median filter to further smooth out the high-frequency noise. The red line in the bottom panel shows the result of this median filtering, and the blue line show the final residual spectrum used for stacking. 

\renewcommand\thefigure{C}
\begin{figure*}[htbp]
    \thinspace\thinspace\thinspace\thinspace\includegraphics[width=1.897\columnwidth]{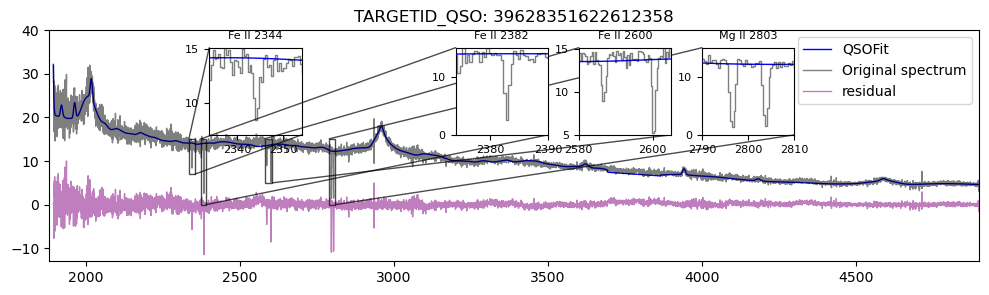}
    \par
    \includegraphics[width=2\columnwidth]{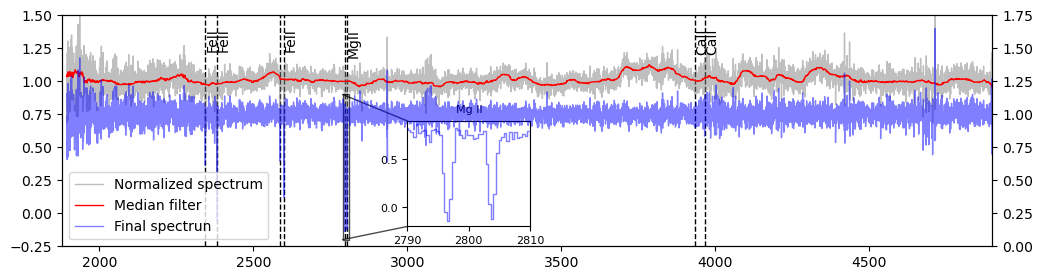}
    \caption{ An example to illustrate how to obtain the final residual spectrum used for stacking. 
    The gray line in the bottom panel represents the normalized spectrum, which is the ratio of the original spectrum to the QSOfit one (black and blue lines in top panel). The residuals between the original spectrum and the QSOfit are represented by the purple line. Some absorption lines (Fe\,\textsc{ii} \& Mg\,\textsc{ii}) and their fits at that location are also zoomed in on. The red line represents the smoothed spectrum with median filter. By further dividing this smoothed spectrum, we obtain the final residual spectrum for stacking (the blue line in the bottom panel). 
    For easy viewing, we shift the blue line downward by 0.25 (see the reference y-axis on the right). }
    \label{fig:spec}
\end{figure*}

\section{The signal-to-noise ratio of co-added spectra}\label{sec:appenC}

We quantify the increase of signal-to-noise ratio ($S/N$) of stacked spectra with increasing the number of spectra used in stacking ($N_{\rm spec}$).
Figure \ref{fig:C1} show the average $S/N$ of single wavelength point in the co-added spetrum as a function of the number of spectra in stacking. 

As shown, the relationship between the $N_{\rm spec}$ and the resulting average $S/N$ is well determined, and it follows a power-law relation: $\log(S/N) = 0.52\log(N_{\rm spec}) + 0.65$. 
This shows that the process of stacking spectra is effective in achieving higher SNR, which is consistent with theoretical expectation: $S/N \propto \ N_{\rm spec}^{0.5}$.

\renewcommand\thefigure{D}
\begin{figure}[htbp]
    \centering
    \includegraphics[width=0.9\columnwidth]{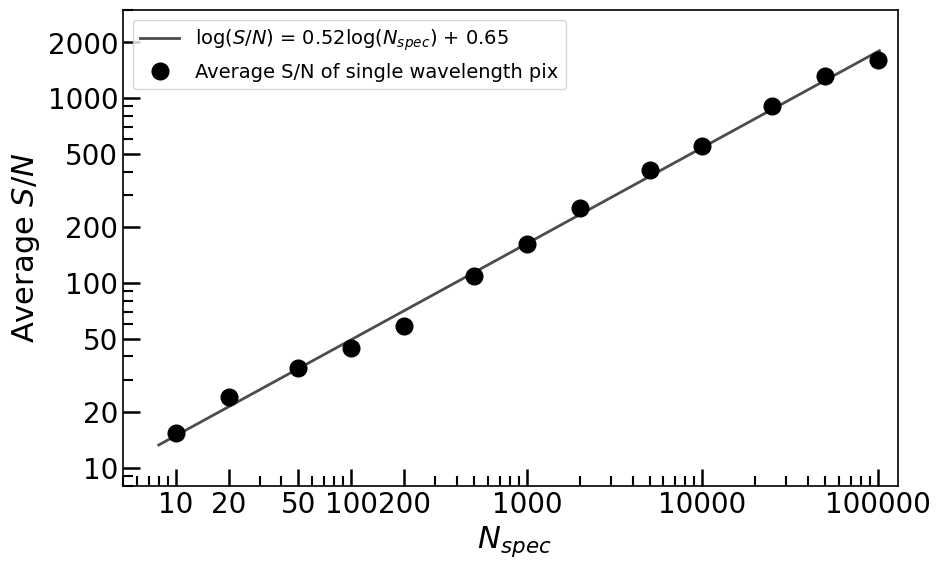}
    \caption{The $S/N$ of co-added spectra as a function of the number of spectra used in stacking. }
    \label{fig:C1}
\end{figure}


\section{Examples of co-added spectra}\label{sec:appenF}
In figure \ref{fig:F1}, we present examples of stacked spectra, and employ a double Gaussian fit for Mg\,\textsc{ii} absorption lines of the stacked spectra. We integrate the wavelength region that located within 3-$\sigma$ of the Gaussian fittings, indicated with the blue shaded region of Figure \ref{fig:F1}. 
For the poorly fitted spectrum (the middle-right panel), we integrate over the fixed range indicated with the gray shaded region.
We show the uncertainties of stacked spectrum at each wavelength point within the fitting range in gray shaded regions. If the absorption signal is buried within the uncertainties of stacked spectra, such as the stacked spectrum in the middle-right panel, we then estimate the upper limit of the equivalent width based on the spectral uncertainties in that range. We examine all the stacked spectra in the present work, and select the stacked spectra without clear Mg\,\textsc{ii} absorption features by eye. 
Due to limitation of the pages, here in Figure \ref{fig:F1} we only show six examples of stacked spectra, and we put all the stacked spectra used in this work on \href{https://docs.google.com/presentation/d/1jF6wtoihM-P7Dg6m74fkX0zbfMV2tt88/edit?usp=share_link&ouid=101552872223175249144&rtpof=true&sd=true}{our Google Drive} for those who are interested in. 
 
\renewcommand\thefigure{E}
\begin{figure*}[htbp]
    \includegraphics[width=1\columnwidth]{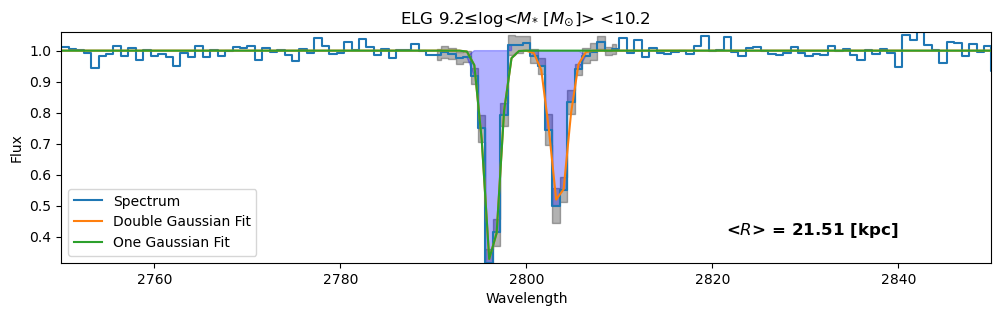}
    \includegraphics[width=1\columnwidth]{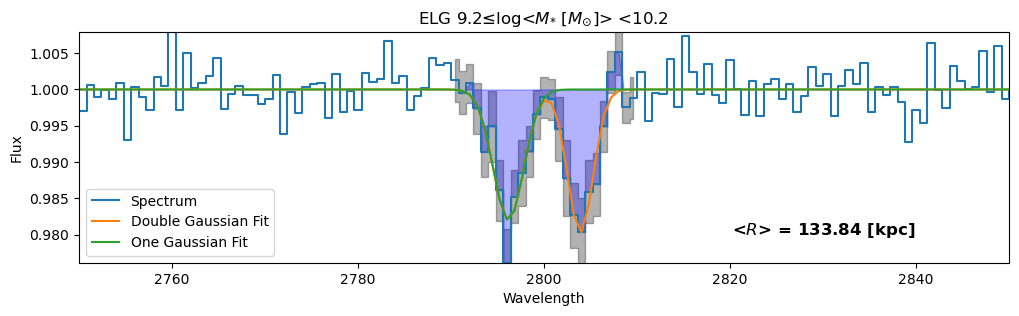}
    \par 
    \includegraphics[width=1\columnwidth]{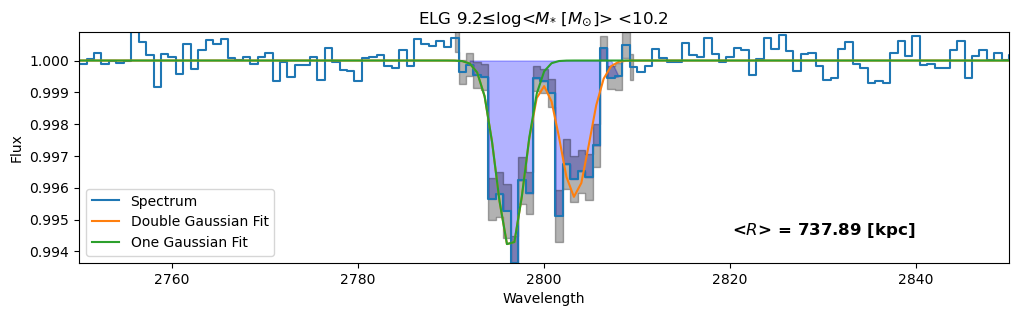}
    \includegraphics[width=1\columnwidth]{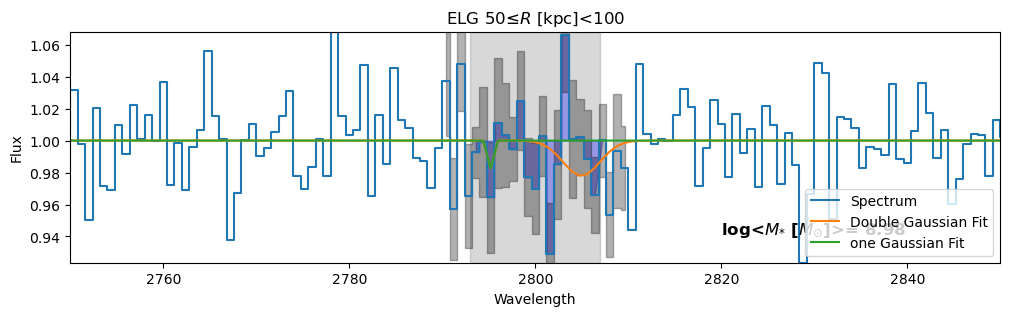}
    \par 
    \includegraphics[width=1\columnwidth]{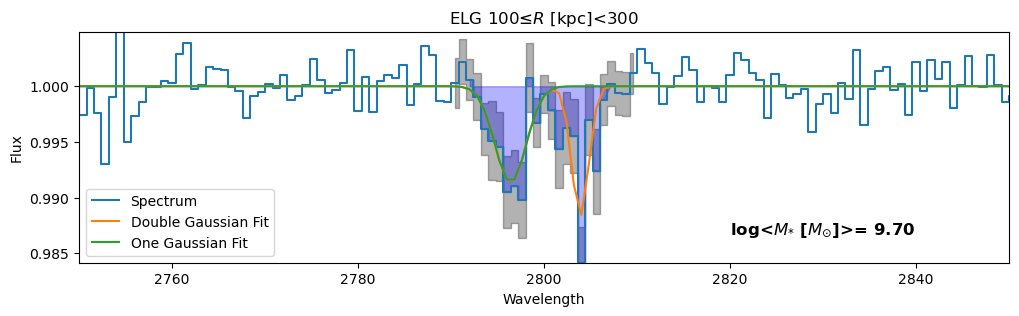}
    \includegraphics[width=1\columnwidth]{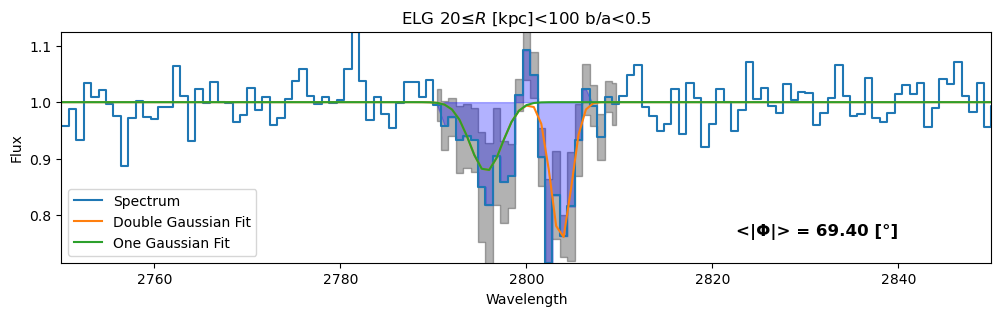}
    \caption{Six examples of the stacked spectra. We adopt a double Gaussian fitting (green \& orange) approach for absorption lines and calculate the integral within a 3-sigma wavelength range as the estimated \rm{EW(Mg\,\textsc{ii})} (blue shaded region, with negative area when flux is greater than 1). For co-added spectra with low signal-to-noise ratio, we employ a fixed wavelength interval for error (black) integration to provide an upper limit estimate for \rm{EW(Mg\,\textsc{ii})}.}
    \label{fig:F1}
\end{figure*}

\section{Validation of our method with eBOSS data} \label{sec:appenB}

In Figure \ref{fig:lrgew}, we find the strength of Mg\,\textsc{ii} absorptions of DESI LRGs are systematically higher than those measured with eBOSS LRGs from \cite{lan18}. Therefore, we propose that this is due to the difference of the LRGs samples, rather than the difference of methods. To examine this, we apply the our method to eBOSS DR17, and obtain the EW(Mg\,\textsc{ii}) for the ELGs and LRGs.  The results are shown in Figure \ref{fig:B1}. For comparison, the results from \cite{lan18} are shown in shaded regions.     

There is a slight difference in the samples of Figure \ref{fig:B1} that we use the data from SDSS DR17, while \cite{lan18} use the DR14. In comparing with \cite{lan18}, here we adopt the measurements of EW(Mg\,\textsc{ii}) using the fitting method (i.e., the first method described in Section \ref{subsec:method}), rather than an integration method. 
Despite this difference, both methods yield near the same results for both ELGs and LRGs when applied to the samples. The good consistency validates the whole processes in our method.  

\renewcommand\thefigure{F}
\begin{figure}[htbp]
    \centering
    \includegraphics[width=0.9\columnwidth]{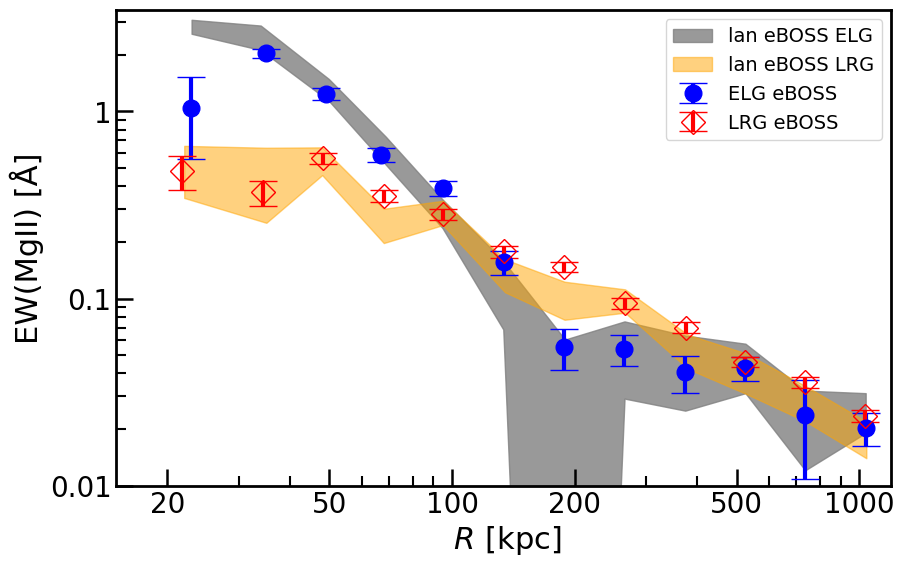}
    \caption{The EW(Mg\,\textsc{ii}) as a function of impact parameter for eBOSS ELGs and LRGs with our method. For comparison, we show the results of \cite{lan18} in the orange and grey shaded regions. }
    \label{fig:B1}
\end{figure}







\end{document}